\documentclass[epjST]{svjour}
\usepackage{graphicx}
\usepackage{amssymb}
\usepackage{amsmath}
\usepackage{color}
\usepackage{hyperref}

\newcommand{\ar}{\arrowvert}
\newcommand{\ra}{\rangle}
\begin{document}
\title{Glueballs as the Ithaca of meson spectroscopy}
\subtitle{From simple theory to challenging detection}
\author{Felipe J. Llanes-Estrada 
\thanks{\email{fllanes@fis.ucm.es}} 
\inst{1} 
}

\institute{Departamento de F\'{\i}sica Te\'orica and IPARCOS, Univ. Complutense de Madrid,
Plaza de las Ciencias 1, 28040 Madrid, Spain}
\abstract{
This compact review about gluonium focuses on a slate of theoretical efforts; 
among the many standing works, I have selected several that are meant to assist 
in the identification, among ordinary mesons, of the few Yang-Mills glueball configurations that populate the energy region below 3 GeV. This includes $J/\psi$ radiative and vector-meson decays, studies of scalar meson mixing, of high-energy cross sections via the Pomeron and the odderon, glueball decays, etc. 
The weight of accumulated evidence seems to support the $f_0(1710)$ as having a large (and the largest) glueball component among the scalars, although no single observable by itself is conclusive. Further tests would be welcome, such as exclusive $f_J$ production at asymptotically high $s$ and $t$. 
No clear experimental candidates for the pseudoscalar or tensor glueball stand out yet, and continuing
investigations trying to sort them out will certainly teach us much more about mesons.
} 
\maketitle

\tableofcontents

\section{Introduction: the glueball as a simple Yang-Mills concept}
\label{intro}

By ``glueballs'' it is broadly understood that we mean the eigenstates of an appropriate Hamiltonian derived from the pure Yang-Mills Lagrangian density,
\begin{equation}\label{YM}
{\mathcal L}_{YM} = - \frac{1}{4} F^a_{\mu\nu} F^{a\mu\nu} \ ,
\end{equation}
with $F^{a}_{\mu\nu}=A^a_{\nu,\mu}-A^a_{\mu,\nu}+igf_{abc} A^b_{\mu} A^c_{\nu}$. If the symmetry group is Abelian, there are no interaction terms (no $f_{abc}$ group structure constants), so that neither photon-photon nor multiphoton states bind. There is no such thing as ``photonballs'' in the absence of matter. 

On the contrary, because the non commutative $SU(3)$ Yang-Mills theory underlying Quantum Chromodynamics is strongly coupled and by all evidence, confining, the (colored) one-gluon states such as
$\int d^3x f(x) A^a(x)|0\rangle$ are not part of its spectrum (they are presumably removed to infinite energy). The spectrum must then be formed of color-singlet two- or multi-gluon states, or glueballs (sometimes ``Gluonium'' is used for the particular case of exactly two gluons, in analogy with $q\bar{q}$ quarkonium).

In conventional lattice gauge theory~\cite{Munster:2000ez}, space-time is rotated to Euclidean four-dimensional space, then discretized at intervals of size $a$, and a change of variables from the Yang-Mills $A^a_\mu$ fields to the parallel-transporter links between two lattice sites, $U(x+a,x)$ is performed. If we could lift the discretization, we could interpret this link variable as a short Wilson line in the lattice direction in which the four-vector $a$ points, 
\begin{equation}
U(x+a,x) = P exp\left( ig\int_x^{x+a} A^b_{\mu}T^b du^\mu \right)
\end{equation}
with $u^\mu \in [0,a^\mu]$ and $T$ the $3\times 3$ color matrix.
Four such links in a closed square of sides $a$ and $b$ of equal length form the gauge-invariant plaquette,
$\tilde{U}_{\mu\nu}(x):= U(x,x+b)  U(x+b,x+a+b) U(x+a+b,x+a)   U(x+a,x)$
from which Wilson's discretized version of Eq.~(\ref{YM}) can be built,
\begin{equation}
\mathcal{L}_{\rm Wilson} = -\frac{2}{g^2} {\rm Re}({\rm Tr}(U)) 
\end{equation}
(The action is obtained by summing over all possible plaquettes, that in the limit $a\to 0$ amounts  to integrating the Euclidean continuation of Eq.~(\ref{YM}).)

The mass of the eigenstates (glueballs) of this discretized theory can, in an unsophisticated analysis, be computed from expectation values of two spatial plaquettes separated by a large time interval $t$,
\begin{equation}
\langle {\rm Tr}(U(t=0)) {\rm Tr}(U(t)) \rangle \propto e^{-m_Gt}\ 
\end{equation}
(which is the Euclidean version of $e^{iHt}$ projected over the lowest eigenvalue, that survives the exponential decay for the longest time). Excited states need to be obtained with smart subtraction of the fundamental one, but this is now routinely done.

The resulting glueball spectrum is obtained as function of the lattice energy scale $a^{-1}$.
To evaluate this, another observable, typically the static potential between color charges, has to be computed and compared with an experimental observable (typically the quarkonium string tension pseudoobservable extracted from spectroscopy with a potential interpretation). There are numerous systematic effects that are addressed in actual lattice computations~\cite{Liu:2000ce}.

An entirely different problem, open to date, is to locate these $G$ states in the physical world where gluons (radiation) are coupled to quarks (matter).

This topical review, that does not intend to be exhaustive nor historical, focuses mostly on that problem. The interested reader can delve into the very extensive literature and
standing reviews of the field~\cite{Crede:2008vw,Mathieu:2008me}. Our purpose here is 
to give a quick topical overview of some selected avenues for glueball identification that we find particularly interesting, promising or classic, presenting alleys of investigation that theorists have suggested. At various points of the article I use results from Effective Lagrangians for hadrons, from the Coulomb-gauge constituent picture, from QCD sum rules, from the flux tube model, or from the AdS-CFT approach.
A quick search of the Inspirehep database reveals that over 1600 scholar articles contain in their titles one of the words ``glueball'', ``gluonium'' or their plurals. I have purposedfully tried to keep the reference list near 100 to contain the review. I have also chosen to focuse on the more contemporary developments (basically, the latest ones come from data taking at BES-III and TOTEM) and, given the nature of this EPJST volume, deemphasize heavier gluonia in the charmonium region in favor of the few glueballs that are lighter than the $J/\psi$.

I have chosen to discuss each of the tree quantum number combinations available for that lightest mass-range, $0^{++}$, $2^{++}$ and $0^{-+}$; because the status of knowledge is different for each of them, and because they may be of interest for different physics phenomena, the review treats them asymmetrically. 
 
Also, for concision, I try not to repeat material: for example, since I discuss the mixing and width of the scalar glueball, I do not cover this for the other two glueballs: because Regge theory is most important for the tensor glueball, I do not discuss the Regge trajectories that may be of interest for the other two; and the same principle applies to the rest of the review.

\section{Pure gauge theory (or quenched approximation)} \label{sec:intro}
\subsection{Lattice spectrum}

Following the lattice computations of the late 90's and early 2000's, most of the community became convinced that the lightest (scalar) glueball was to be searched for among the $f_0$ mesons in the 1.3-2 GeV region~\footnote{
There is a minority view that the $\sigma$-meson has a Fock-space component of the lowest scalar gluonium as hinted by early bag-model computations and more elaborate QCD sum rules.
The approach accommodates its large coupling to $\pi^+\pi^-$ and to (subthreshold) $K^+K^-$ by invoking a large violation of the OZI rule at these lowest energies. 
(By contrast, good satisfaction of the OZI rule 
in the $\sim 1.7$ GeV energy region suggests sizeable couplings to  $\eta^{(')}\eta^{(')}$ pairs with large glue content.) }

\begin{figure}
\includegraphics[width=\columnwidth]{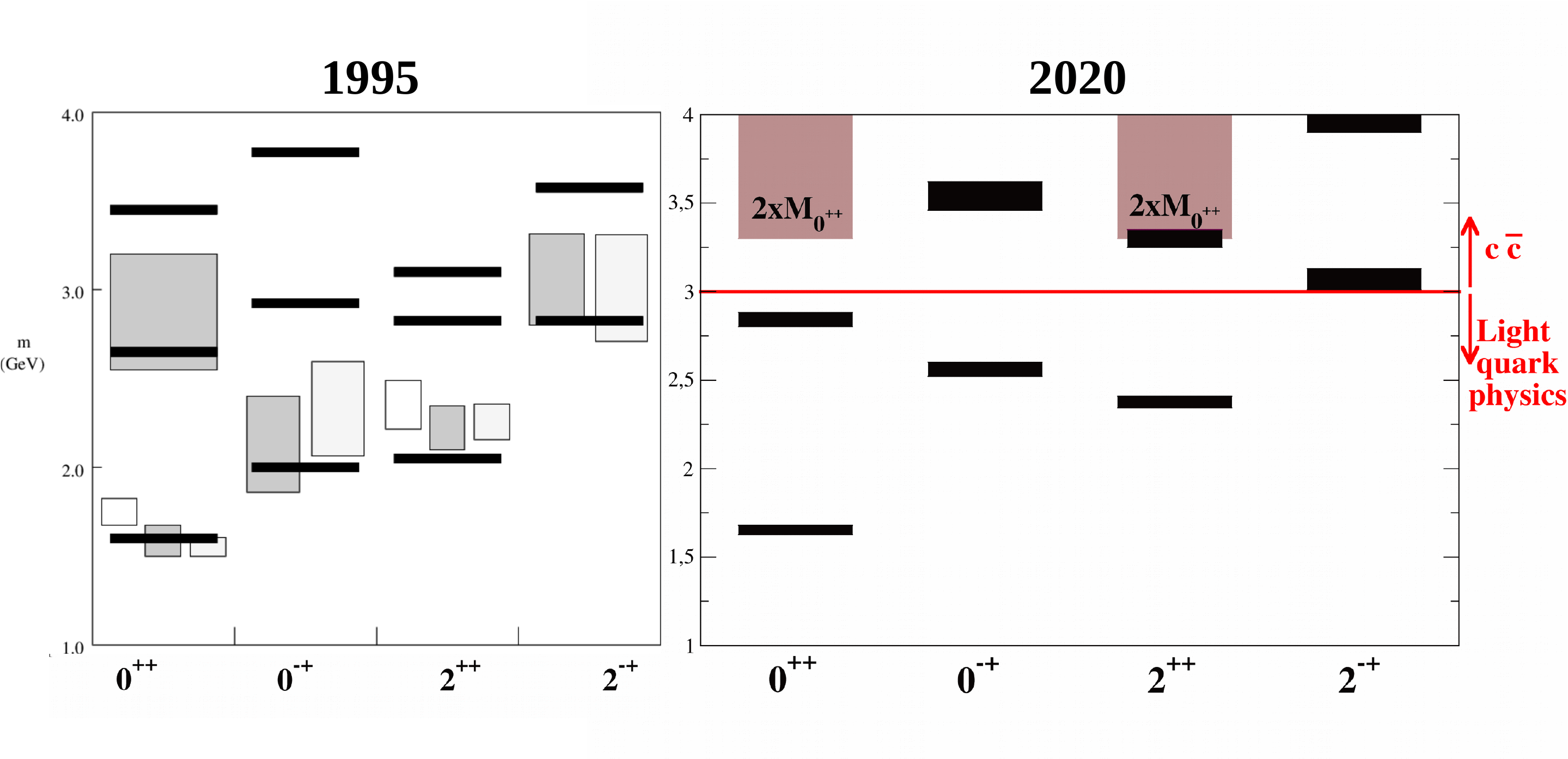}
\caption{\label{fig:glueballspectra} Change of the computed glueball spectrum in 25 years. Left (from~\cite{Szczepaniak:1995cw}, with APS permission): the boxes were the lattice computations at the time, whereas the narrow black lines stand for the NCSU Coulomb-gauge BCS+Tamm-Dancoff model calculation. 
Right: the most recent lattice computation~\cite{Athenodorou:2020ani} (black lines) now has much reduced uncertainties. The qualitative comparison of the spectra is reasonable. I have marked, in the right graph, the divide between charmonium and light-quark spectroscopy, as well as the two-glueball continuum of pure YM theory. 
}
\end{figure}

Figure~\ref{fig:glueballspectra} shows the evolution of the $C$-even (two-gluon like) spectrum in the last twenty-five years. Around 1995 the lattice gauge theory prediction was quite uncertain (see the width of the boxes in the left plot; the lines come from the model approach in the next subsection~\ref{subsec:constgluons}, the NCSU Coulomb-gauge Hamiltonian) but it has become quite accurate with the years, as seen in the right plot.

We can, with quite some certainty, state that the glueballs expected below 3 GeV have the $J^{PC}$ quantum numbers of the $f_0$ family ($0^{++}$), the $f_2$ family ($2^{++}$) and the $\eta$ one ($0^{-+}$). None of these states is faneroexotic (manifestly exotic), instead having conventional $q\overline{q}$ quantum numbers. 

In the charmonium region and above there can be exotic-quantum number~\cite{Meyer:2010ku} glueballs, but they will compete with hybrid $q\overline{q}g$ mesons~\cite{LlanesEstrada:2000hj,Soto:2017one}, tetraquarks and others.

Since this topical review is intended for a volume dedicated to light quark physics, most of the discussion will concern the $f_0$, $f_2$ and $\eta$--like glueballs.

\subsection{The gluon constituent picture} \label{subsec:constgluons}

It is often stated that gluons are massless particles, and must be so because of gauge symmetry. 
This affirmation is based on the lack of gauge invariance of a Proca-like Lagrangian density, that adds a term to Eq.~(\ref{YM}),
\begin{equation}
L_M = \frac{m_g^2}{2} A^{a\mu} A^a_\mu\ .
\end{equation}
In this sense, yes, classical Yang-Mills theory cannot accommodate a gluon mass. 

Yet it is obvious that the gluon degree of freedom is dynamically gapped because of the partly discrete nature of the hadron spectrum. If adding a massless gluon with $J^{PC}=1^{--}$  did not cost any energy, one could construct baryons of arbitrary quantum numbers with the same $940$ MeV mass of the proton! That is obviously not the case, with the lowest proton excitation being the $\Delta(1232)$.

Thus, gluons need to satisfy a gapped dispersion relation brought about by the interaction terms in the quantum theory~\cite{Cornwall:1982zn} (and this leads directly to a discrete glueball spectrum). Examples of the phenomenon are easily borrowed from electrodynamics,
\begin{equation}
\omega(k)^2 = k^2 + m_g^2 
\end{equation}
with $m_g$ stemming from a plasma cutoff frequency in a conductive medium (this, in QCD, is deployed in heavy-ion collision studies, with $m_g^2 \propto \alpha_s T^2$ at finite temperature, see for example~\cite{Alam:1996wp}); or with $m_g$ arising from boundary conditions such as in a microwave cavity, which is deployed in the bag model of hadrons, also used to compute glueball spectra~\cite{Jezabek:1982ic}. For example, the Transverse Electric modes in a bag of radius $R$ have an energy given by
\begin{equation}
\tan (\omega R) = \frac{\omega R}{1-\omega^2 R^2}\ ,
\end{equation}
with lowest mode (``mass'') equal to $\omega\simeq 2.74/R$ (and $4.5/R$ for a TM mode)~\cite{Lagerkvist}.
Of course, the bag model as other approaches containing hadron-external condensates must face the inconvenience of the cosmological constant~\cite{Brodsky:2012ku}.
Another well-known such example is the Higgs mechanism in which an additional field is used to break a global symmetry, with the resulting Goldstone bosons providing the longitudinal modes of the electroweak $W$ and $Z$ bosons, and their mass being given by the Higgs condensate.

But in the context of Quantum Chromodynamics, the most popular approaches to describe the mass gap are based on many body approximations to the strongly coupled gauged problem itself.
For example, a Coulomb-gauge gap equation based on the confining Coulomb potential between gluons~\cite{Szczepaniak:1995cw},
\begin{equation}
\omega_k^2 = k^2 +\frac{N_c}{4} \int \frac{d^3q}{(2\pi)^3} V_{\rm Coulomb}({\bf k}+{\bf q})
(1+ (\hat{\bf k}\cdot \hat{\bf q})^2) \frac{\omega_k^2-\omega_q^2}{\omega_k} 
\end{equation}
provides a running gluon energy $\omega_k \simeq \sqrt{k^2 + m_g^2 e^{-(k/\kappa)^2}}$ and a canonically transformed vacuum/ground state $\ar 0\rangle \to |ar \Omega \rangle$ that approximates the exact QCD one.

Instead of a gluon dispersion relation approaching a masslike constant
at vanishing momentum, other authors employ one where the gluon ``mass'' diverges
in the infrared, as variationally estimated by Feuchter and Reinhardt~\cite{Feuchter:2004mk}
in accordance with Gribov-Zwanziger's confinement scenario,
\begin{equation} \label{disprel2}
\omega(k) = \sqrt{k^2 + \frac{M^4}{k^2}}
\end{equation}
where $M\simeq 880 $ MeV also reproduces lattice glueball spectroscopy.

This hadron rest frame picture has been, with quite some labour, been extended to the covariant Dyson-Schwinger+Bethe-Salpeter approach in Landau gauge~\cite{Meyers:2012ka,Huber:2020ngt,Sanchis-Alepuz:2015hma,Souza:2019ylx,Kaptari:2020qlt}.

Yet an advantage of the Hamiltonian Coulomb gauge formulation is that the absence of a $J=1$ glueball in the low-lying spectrum is immediate to understand: Yang's theorem~\footnote{A recent well known application thereof was to exclude $J=1$ for the Higgs boson, as its decay $h\to \gamma\gamma$ was quickly identified.} states~\cite{Lee:1981mf} that two identical transverse bosons of spin 1 each cannot couple to total $J=1$. 
Thus, if the low-lying glueball spectrum is dominated by $\ar gg\rangle$ states in the Coulomb gauge formulation where by construction $\nabla\cdot {\bf A}=0$, so that transversality is guaranteed,
a spin-1 glueball is forbidden. 
This is by no means automatic in covariant approaches, such as the Landau gauge Bethe-Salpeter formulation
in which $\partial_\mu A^\mu = 0$ is not sufficient to implement Yang's theorem. A detailed dynamical mechanism must then be responsible for removing the $J=1$ glueball. Likewise, in the AdS-CFT approach to glueballs (that are thought to arise from a supergraviton spectrum in a theory dual to QCD)
a light spin-1 glueball appears~\cite{Vento:2017ice} alongside the $0^{++}$ and $2^{++}$, though strong splitting, for reasons not totally clear to me, can raise the state with spin 1 to higher mass~\cite{Brower:2000rp}.
The same inconvenience is present in constituent approaches in which the constituent gluons are treated as massive  Proca spin-1 bosons: it is not easy to get rid of the $J=1$ glueball~\cite{Bicudo:2004tx,Mathieu:2008me}. Thus, the Coulomb-gauge dynamical mass generation picture remains a competitive contender to understand the low-mass lattice glueball calculation.

\newpage

\section{Coupling to quarks and glueball width}

The lattice glueball spectrum has also been looked at with unquenched QCD that includes dynamical quarks, for example in~\cite{Gregory:2012hu}. This group finds that the effect of including quarks in the simulation is to raise the masses of all the states, even up to 30\%. The scalar glueball is only lifted by 5\%, from the $1.71-1.73$ GeV of other calculations up to $1.8(6)$. Other computations cited therein, however, see the scalar glueball mass descending. 
Ultimately, in a full QCD calculation, all scalar $f^i_0$ mesons give a signal when computing scalar-scalar correlators, unless the matrix element $\langle \Omega \arrowvert \mathcal{O}_s \ar f_0^i\rangle$ exactly vanishes, which is not to be expected in a theory of the strong interactions. 
One can speculate that this would be an explanation for the instability seen in such calculations. Ultimately, there is no such thing as  ``unquenched glueballs'', at that point one is simply computing the full scalar meson spectrum. 

One thing that can be done, however, is to adiabatically track the fate of the pure Yang-Mills glueball pole as the coupling to quarks is slowly turned on. To my knowledge, such calculation has not been carried out. The most interesting quantity that would come out of it would be a nonperturbative computation of the ``glueball'' width (at the end point, one of the $f_0$s). Naturalness suggests that $\Gamma_G\sim \Delta M_G$ (the real and imaginary part of the glueball mass acquire contributions of the same order upon unquenching), so that $\Gamma~O(0.1)$ GeV is conceivable.

The QCD sum rules approach employs a dispersive analysis with simple model elements 
to extract the glueball width. A standard analysis~\cite{Shuiguo:2010ak} would proceed by
modeling the spectral function of QCD in the scalar channel 
\begin{equation}
\Pi(q^2=s)=\int d^4x e^{iq\cdot x} \langle \Omega \ar T \mathcal{O}_{\rm scalar} (x) \mathcal{O}_{\rm scalar} (0) \ar \Omega \rangle 
\end{equation}
as
\begin{eqnarray}
{\rm Im} \Pi (s) &=& \rho^{\rm had}(s) + {\rm Im} \Pi^{\rm pQCD}(s) \theta(s-s_0) \nonumber \\
&=& \sum_i^{\rm res} \frac{f_i^6 m_i \Gamma_i}{(s-m_i)^2+\Gamma_i^2/4 + m_i^2 \Gamma_i^2} + {\rm Im} \Pi^{\rm pQCD}(s)\theta(s-s_0) 
\end{eqnarray}
with $f_i=\lambda^i_0 s \ \theta(m_\pi^2-s) + (\lambda^i_0 m_\pi^2 + \lambda_1^{i3})\theta(s-m_\pi^2)$
carrying a couple of fittable strength constants $\lambda^i_0$ and $\lambda^i_1$ to model the coupling of the QCD current to that hadron state, $f_i= \langle \Omega \ar \mathcal{O}_{\rm scalar} \ar f_0^i\rangle $, and the pQCD part computed in perturbation theory.  
This very rough model (note the Breit-Wigner approximation to the scalar mesons!) of the physical 
spectral function is then related via a dispersion relation to a spacelike-$q^2$ computation carried out in pQCD together with a classical instanton background. When the dust settles, a glueball width is extracted from the corresponding parameter $\Gamma_i$, that I elevate to table~\ref{tab:width}. The coupling to quarks is computed in perturbation theory through the pQCD elements, but this is a different approximation from the others here discussed, because that part of the computation takes place for unphysical or very large $q^2$, not in the soft hadron region.

A typical constituent-like computation of the glueball width into two mesons would proceed by evaluating the coupling in second-order perturbation theory
\begin{equation}
g=\sum\int \langle \psi_G^* gg \ar  H_{\rm int}\ar q\bar{q}g\rangle \frac{1}{M_G-E_{q\bar{q}g}}\langle q\bar{q} g \ar H_{\rm int} \ar q\bar{q}q\bar{q}\rangle
\end{equation}
through all intermediate state that are hybrid mesons (the Tamm-Dancoff approximation glueball wavefunction $\psi_G^*$, as well as all the masses, need to be calculated ahead before the Feynman diagrams in $H_{\rm int}$ are included). Such approximation is supposed to work better (but yield broader glueballs) the higher the mass $M_G$, because (a) the coupling constant $\alpha_s$ becomes smaller with increasing gluon momentum so that perturbation theory is sounder, and (b) there are more abundant intermediate hybrid mesons in the high spectrum, so some will always be near the energy-shell $M_G$ in the decay. This was estimated for the scalar glueball at $M_G\simeq 1.8$ GeV~\cite{Bicudo:2006sd} and found to yield a relatively narrow state with $\Gamma_G\simeq 0.1$ GeV, with a larger $\pi\pi$ than $K\bar{K}$ component as demanded by phase space, as shown in table~\ref{tab:width}.

The best known lattice computation~\cite{Sexton:1996ed}, in quenched approximation, proceeded by matching a three-point function between the scalar glueball current and two pseudoscalar currents $\bar{\psi} \gamma_5 \psi$. It found a $\sim 1.7-1.8$ GeV glueball, of narrow width $\Gamma_G=0.108(29)$ GeV, and interestingly,
seemingly asymmetric couplings favoring decays through the strange quark; this topic will be picked up again in subsection~\ref{sratherthanu} below.

\begin{table}
\caption{\label{tab:width}
Theory estimates of the $f_0$-like scalar glueball width for approaches that place it in the $1.5-1.75$ GeV mass region, and experimental estimates of the scalar meson widths in the 1-2 GeV interval. The lattice and semiperturbative Coulomb model estimates include only two body ($\pi\pi$, $K\bar{K}$, etc.) decays, so they are lower bounds to the total width. Overall, a narrow glueball with $\Gamma_G\sim 0.2$ GeV seems a plausible theory prediction (I do not list the additional 1.81 GeV structure in $\omega\phi$ since later analysis confirmed that a new resonance should also be manifest in $K\bar{K}$, which it is not, and that it likely is the same $f_0(1710)$ seen at a higher mass due to the $\omega\phi$ threshold distortion~\cite{Wang:2011tm,MartinezTorres:2012du}).}
\begin{tabular}{|c|ccccc|} \hline
Method & Sum rules & Lattice (quenched) & Coulomb-$gg$ & $G$-dominance & Flux-tube \\ \hline
$\Gamma$ (GeV) & ${\bf 0.23(13)}$\cite{Shuiguo:2010ak} & {\bf 0.11(3)}~\cite{Sexton:1996ed} & {\bf 0.1}~\cite{Bicudo:2006sd} & $>${\bf 0.25-0.39}\cite{Burakovsky:1998zg} &  ${\bf \sim 0.18}$~\cite{Iwasaki:2003cr}  \\ \hline\hline
Meson~\cite{Zyla:2020zbs}   & $f_0(1370)$ & $f_0(1500)$ & $f_0(1710)$ & $f_0(2020)$? & \\ \hline
$\Gamma_{\rm exp}$ &   {\bf 0.2-0.5}          & {\bf 0.11(1)}   & {\bf 0.12(2)} & $\sim{\bf 0.4}$ & \\ \hline
\end{tabular}
\end{table}

Not all approaches yield such narrow glueballs. Among standing calculations for a broad scalar glueball,
I highlight a model computation~\cite{Burakovsky:1998zg} that employs a so called ``glueball dominance hypothesis'' to reduce the parameter space of a mixing calculation of $G$, $s\bar{s}$ and $q\bar{q}$ light quarkonium at the level of the meson mass matrices. Their characteristic hypothesis is that the different flavors of scalar quarkonium are not connected directly, but mix only through an intermediate glueball state, as inspired by large-$N_c$ ideas. The authors are also inspired by the flux tube and $^3P_0$ decay model. They do assume flavor-blind couplings, and uncharacteristically, find a broad scalar glueball with $\Gamma=0.25$ GeV at least, and even above $0.39$ GeV. This is driven by the decay $f_0\to a_1\pi$ that accounts for half the width and is a dominant decay mode.

Though these authors place the dominantly glueball-state mass just above 1.7 GeV and the $f_0(1710)$ is their prefered candidate, neither the width of this meson as later measured matches their expectations, nor has the $a_1\pi$ decay mode been listed yet.

Flux-tube breaking arguments with $\Gamma_G \propto M_G$~\cite{Iwasaki:2003cr} naturally suggest that excited glueballs will be broader, in line with other types of mesons.

\subsection{Exploiting symmetry in glueball decay and mixing} \label{subsec:symmetrymixing}
Glueballs are much heavier than pseudoscalar and vector mesons, entailing several possible open strong decay channels. It is obvious that their decays are important to identify them, 
and this section therefore addresses some of them.

Several groups~\cite{Rosenzweig:1981cu,Cheng:2006hu,McNeile:2000xx,Narison:1996fm}  have  addressed~ the configuration mixing of glueballs with other ordinary or exotic mesons.
It is clearly necessary to have criteria which bear on the two topics of glueball identification and mixing, but also to be able to theoretically define that mixing. 

The Coulomb gauge QCD formulation offers a full Fock expansion of a meson that includes only quarks and (``physical'') transverse gluons, schematically
\begin{equation} \label{Fock}
\ar M \rangle = \sum\int \left(
\alpha_1 \ar q\bar{q} \rangle + \alpha_2 \ar gg \rangle   + \alpha_3 \ar q\bar{q}g \rangle + \alpha_4 \ar q\bar{q} q\bar{q} \rangle + \alpha_5 \ar ggg \rangle + \dots
\right) \ .
\end{equation}
With a well-defined canonical transformation~\cite{LlanesEstrada:1999uh} one can choose $g$ and $q$ to correspond to the current fields in the free Lagrangian, or rotated fields whose quanta are massive-like constituents due to the interactions. 

The inconvenient of this intuitive expansion is the difficulty to experimentally access it because of its frame (and gauge) dependence: the similar light-front gauge expansion useful in subsec.~\ref{scalartensor} below will have different $\alpha_i$ coefficients.

Either of them could in principle be accessed by adequately projecting lattice correlators, but this has not been performed. What lattice can more easily provide is a proxy to that expansion, the relative strengths with which different composite field operators couple $\ar M\rangle$ and the vacuum $\ar \Omega \rangle$. This has the inconvenience of including longitudinal gauge modes/scalar potentials, and components of different representations of the rotation group packed inside the representations of the Lorentz group and its lattice symmetry reduction. 

Because of the difficulty, other methods have been devised. One is to extract gauge-independent content from the large-$N_c$ expansion around $N_c=3$~\cite{Cohen:2014vta}. While interesting, one issue there is that large-$N_c$ only sorts wavefunction configurations into classes: for example, 
both conventional $q\bar{q}$ and hybrid mesons have widths 
$\Gamma_{q\bar{q}} \propto \frac{1}{N_c}\propto \Gamma_{q\bar{q}g} $, so they cannot be distinguished~\footnote{A further ambiguity is in the definition of a tetraquark: how does $q\bar{q}q\bar{q}$ with $2=3-1$ pairs generalize to more than three colors, as 2  or as $N_c-1$ pairs? The ambiguity is resolved in~~\cite{Cohen:2014vta}.}.
In the end, glueballs are expected to be narrower, $M_G\propto 1$, $\Gamma_{G} \propto \frac{1}{N_c^2}$ instead, so they can be separated with lattice data for different $N_c$ values. But concerning the physical world, the only statement is that glueballs are qualitatively narrower than conventional mesons. 

Finally, effective hadron models such as shown in subsection~\ref{subsubsec:mix} study the mixing of an additional singlet particle to which some additional ``glueball''-like dynamics is adscribed based on underlying physics, and it is in this sense that most mixing analysis are presented. The connection of the information gained to the microscopic expansion such as Eq.~(\ref{Fock}) is contained in that dynamical statement only.

\subsubsection{Flavor-blind quark-gluon vertex}

The QCD Lagrangian features a flavor $SU(3)$ symmetric quark-gluon vertex: all flavors  equally couple to the gluon. This has been a motivation to write flavor-symmetric chiral Lagrangians, such as used in many mixing analysis, some examples being recalled in the next subsection~\ref{subsubsec:mix}.

If glueballs decay/rehadronize via a chain $gg\to gq\bar{q}\to q\bar{q}q\bar{q} \to MM$, which is disputed~\cite{Chao:2005si}, the strong dynamics is not bound to disrupt flavor symmetry much, and for example, its coupling to $\pi\pi$ is expected to be similar to that to $K\bar{K}$. After accounting for phase space, a 1.7-1.8 GeV glueball would have a width around 0.1 GeV and $\pi\pi$ would be dominant~\cite{Bicudo:2006sd}. This flavor symmetry in the couplings is expected for most glueballs in any case, but much of the analysis in the next subsection~\ref{subsubsec:mix} assumes that it particularly applies to the scalar glueball.

On the contrary, should the dominant decay mode be $gg-q\bar{q}$ mixing, chiral symmetry is more important for the scalar glueball, badly breaking flavor symmetry; this is quickly overviewed in subsection~\ref{sratherthanu}.

\subsubsection{Exploiting flavor symmetry in a mixing analysis} \label{subsubsec:mix}

A very well known 1995 analysis of Crystall Ball data by Amsler and Close~\cite{Amsler:1995td}, 
among other works, gave support to the hypothesis that $f_0(1500)$ was largely the $0^{++}$ glueball $G$; 
this is therein introduced as an additional singlet state, coupling to the two-pseudoscalar meson pairs according to 
\begin{eqnarray}
\langle G \ar H_{\rm int} \ar \pi\pi \rangle =1 &\phantom{multimixing} & \langle G \ar H_{\rm int} \ar K\bar{K} \rangle :=R \nonumber \\
\langle G \ar H_{\rm int} \ar \eta\eta \rangle =\frac{1+R^2}{2} &\phantom{multimixing} & 
\langle G \ar H_{\rm int} \ar \eta\eta' \rangle =\frac{1-R^2}{2}
\end{eqnarray}
with the limit of exact flavor $SU(3)$ symmetry reached by setting $R=1$ and, after accounting for the charge multiplicity, leads to decay proportions 
\begin{equation}
G\to \pi\pi: \eta\eta : \eta \eta': K\bar{K} = 3:1:0:4 \ .
\end{equation}
(An accurate prediction would additionally need to account for the difference in phase space.) 
The authors then concluded that $f_0(1500)$ had decay features consistent with the glueball assignment,
though a small proportion of this singlet would also be mixed in the $f_0(1370)$.

\begin{figure}\begin{center}
\includegraphics[width=0.4\columnwidth]{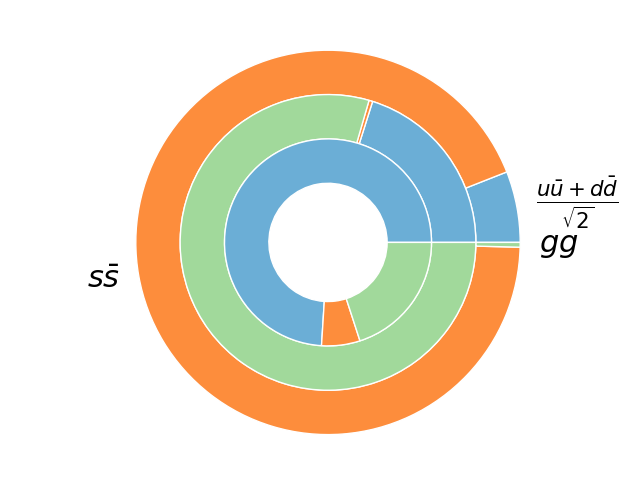}\hspace{-0.4cm}
\includegraphics[width=0.4\columnwidth]{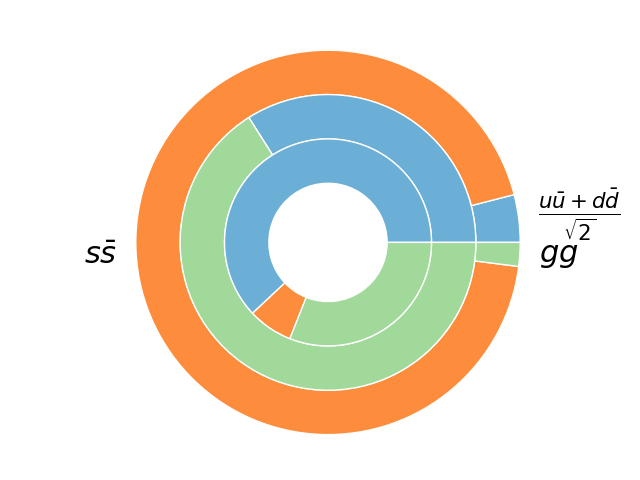}
\includegraphics[width=0.4\columnwidth]{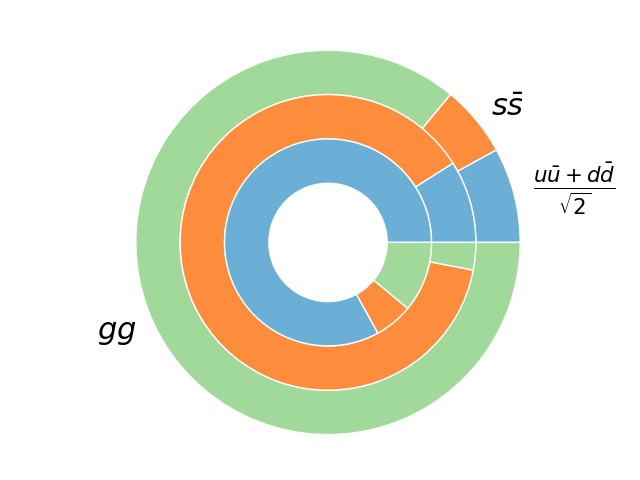} \end{center}
\caption{\label{fig:mixing} Example computations of glueball-like
and quarkonium-like mixing. From inner to outer rings, the composition of the
$f_0(1370)$, $f_0(1500)$ and $f_0(1710)$ is given. Proceeding counterclockwise from the
$OX$ axis, the slices correspond to $\frac{u\bar{u}+d\bar{d}}{\sqrt{2}}$, $s\bar{s}$ and
the glueball. The top plots correspond to the first and third solutions, respectively,
of Giacosa {\it et al.}~\cite{Giacosa:2005zt}, while the bottom plot shows the mixing 
resulting from a glueball-as-dilaton chiral model~\cite{Janowski:2014ppa} }
\end{figure}

More sophisticated analysis in the next two decades proceeded by constructing full chiral Lagrangians including the additional glueball-singlet state. Among the many studies I have selected two representative ones~\cite{Giacosa:2005zt,Janowski:2014ppa} whose outcomes are shown in figure~\ref{fig:mixing}, including the three scalar states $f_0(1370)$, $f_0(1500)$ and $f_0(1710)$, presumed a mixture of three particles with flavor couplings characteristic of $\frac{u\bar{u}+d\bar{d}}{\sqrt{2}}$, $s\bar{s}$ and a singlet $G$ presumed to be the glueball.

The top plots of figure~\ref{fig:mixing}, produced with data from~\cite{Giacosa:2005zt}, suggested that indeed most of the glueball is spanning the state $f_0(1500)$ as also suggested by Amsler and Close. The difference is that, while the left top plot assumes that the direct couplings $G\to\pi\pi,K\bar{K}$ are suppressed and $0^-0^-$ glueball decay proceeds by mixing with conventional quarkonium (exactly the opposite case will be discussed in subsection~\ref{sratherthanu} below), the right plot allows for direct decay. In the later case, some of the glueball component shifted to the lightest $f_0(1370)$. 

Other analysis with similar flavor symmetry content and experimental data offer a quite different picture, such as that from~\cite{Janowski:2014ppa} that assigns most of the glueball to the $f_0(1710)$ (bottom plot in figure~\ref{fig:mixing}). 

Conventional mesons are interpreted in the context of a linear sigma model (a specific realization of chiral dynamics less general than Chiral Perturbation Theory) to reduce parameter space, with $q\bar{q}, s\bar{s}\sim \sigma_i$, and the $U(3)_L\times U(3)_R$ chiral invariant effective Lagrangian being constructed from a field multiplet that incorporates these scalar and the pseudoscalar mesons, $\Phi =\sum(S_i+iP_i)\frac{\lambda_i^{\rm Gell-Mann}}{2}$.
In that model, the extra glueball state is not only assumed to be a flavor singlet, but endowed with additional dynamics stemming from the assumption that it reflects the loss of dilatation symmetry of the Yang-Mills Lagrangian in Eq.~(\ref{YM}).
This is implemented by introducing an auxiliary effective dilaton field $G$ with Lagrangian
\begin{equation}
{\mathcal{L}}_{\rm dilaton} = \frac{1}{2} (\partial_\mu G)^2 - \frac{m_G^2}{\Lambda^2} 
\left( \ln \left( \frac{G}{\Lambda}\right) -\frac{1}{4}\right)\frac{G^4}{4}
\end{equation}
with minimum at $\langle G \rangle =\Lambda$ and particle excitation above it with mass $m_G$. If the glueball/dilaton is further assumed to saturate the trace of the dilatation current brought about by quantum effects (trace anomaly), the authors obtain a relation between $\Lambda$ and $m_G$ that become interdependent. For a ``narrow'' particle-like glueball in the 1.5-1.7 GeV energy range, $\Lambda\sim 3$ GeV, whereas for a more reasonable $\Lambda\sim 0.4$ GeV in the hadronic regime, the glueball becomes a very broad structure.

In the first case, the pattern of decays of the scalar mesons is best fit if the mixing angles (that are in these approaches independent model parameters) are as in the bottom plot of figure~\ref{fig:mixing}, with the $f_0(1710)$ predominantly the glueball.
In the second case, at odds with the large $N_c$ expectation, my interpretation is that we would think of the glueball as a background, and the glueball would not correspond to any of the experimentally studied $f_0$ mesons.

\subsubsection{Flavor-symmetry breaking decay of the scalar glueball} \label{sratherthanu}

Building on earlier work, Chanowitz~\cite{Chanowitz:2005du} conjectured, on the basis of an all--orders perturbative QCD computation, that the scalar glueball couples more strongly to $K\bar{K}$ than $\pi\pi$ (as suggested by suppression of its coupling to $q-\bar{q}$ being proportional to $m_q$). 
The argument rests on conservation of chirality by QCD without quark masses: then, the only appearance of the quark spinor in the Lagrangian is $\bar{\psi}_L \gamma^\mu T^a \psi_L A_a + L\to R$.

 When the two gluons annihilate into two quarks (thus, the matrix element corresponds to gluonium/quarkonium mixing),
the created quark and antiquark have the same chirality at all orders of perturbation theory (since iterating the $L-L$ vertex just written never changes $L$ to $R$, for example). Chirality and helicity coincide for the quark, but are opposite for the antiquark, so they appear with opposite helicities. Now, since in the rest frame the momenta are opposite, ${\bf p}_{\bar{q}}=-{\bf p}_q$, ${\bf S}_q\cdot{\bf p}_q=
-{\bf S}_{\bar{q}}\cdot{\bf p}_{\bar{q}}$ (opposite helicities) implies that the spin projections over a fixed $OZ$ axis are actually the same, so that $S^z_{q+\bar{q}}=\pm 1$. This $S=1$ is actually fine to yield a $0^{++}$ quarkonium with $S_{q\bar{q}}=1$, the problem is that the necessary $L_{q\bar{q}}=1$ cannot be reached from an $S$-wave gluon-gluon wavefunction (the angular integral vanishes).

At order $m_q$ however, the scalar term $m_q \bar{\psi}\psi$ violates chiral symmetry and allows for an $L\cdot S$ coupling providing extra orbital angular momentum.

Comparing this QCD theory input with meson analysis, Albaladejo and Oller~\cite{Albaladejo:2008qa} favor the  $f_0(1710)$ scalar as having a larger gluonium component. This is natural given their finding that 
$\Gamma_{\pi\pi}/\Gamma_{K\bar{K}}\simeq 0.32(14)$: the coupling of this meson is larger to $K\bar{K}$ than $\pi\pi$, as can be seen comparing the first and second plots from the top in figure~\ref{fig:MMchannels} that will be discussed later on.

These authors also find that a pole at around 1.6 GeV and somewhat influencing $f_0(1500)$, behaves as a glueball, which is quite surprising since the first excited scalar glueball is not expected below 2.5 GeV (see figure~\ref{fig:glueballspectra}). The explanation is that this pole comes from the $\eta \eta' $ coupled channel and would never be seen in a quenched lattice calculation.

What the all--orders perturbative QCD argument of~\cite{Chanowitz:2005du} really suggests is that $gg$-$q\bar{q}$ mixing is suppressed by $m_q$, which would naturally explain the small amount of $q\bar{q}$ quarkonium found in some analysis such as in the bottom plot of figure~\ref{fig:mixing}; that this mixing dominates the decay is then on a less solid basis, since as already mentioned, the decay might proceed by $q\bar{q}q\bar{q}$ intermediate states that easily hadronize into two mesons by ``fall-apart'' decay.

As a final remark let me note that dynamical symmetry breaking trascends an all--orders computation and requires an infinite resummation, for example in the form of a Dyson-Schwinger equation. Still, because the typical momentum of a constituent--like gluon in a glueball is of order $M/2$, the running quark mass has dropped sufficiently by that scale (many hundreds of MeV) that chiral symmetry is a reasonable approximation, with $m_u\sim m_d$ plausibly in the 10-20 MeV range or so, already small enough for Chanowitz's argument to make sense.

\subsubsection{The axial anomaly and the pseudoscalar glueball} \label{subsubsec:anomaly}

One sometimes reads that the glueball-quarkonium mixing in the pseudoscalar channel is responsible 
for raising the mass of $\eta_{\rm singlet}$  (in turn, a mixture of the physical $\eta$ and $\eta'$ mesons) respect to a reference level in Gell-Mann's octet. This must be incorrect since 
the variational principle, a simple theorem of linear algebra, guarantees that the mixing of two states \emph{lowers} the mass of the lightest one while raising that of the heaviest (``level repulsion'' in many-body jargon). 

Thus, the supposed mixing of the $\eta/\eta'$ system and the pseudoscalar glueball is not the cause of the excess mass in that system. That mixing is, to date, unknown. But the large difference in masses
($m_\eta =547$ MeV, $m_\eta' = 958$ MeV, $m_{0^{-+}G}>2$ GeV) suggests that the mixing might not be a dominant feature.

What is true is that the anomalous term in the axial current of QCD,
\begin{equation}
\partial_\mu J^\mu_5 = \frac{3\alpha_s}{4\pi} F^{\mu\nu} \tilde{F}_{\mu\nu} \ (\equiv \partial_\mu K^\mu)\ .
\end{equation}
with $\tilde{F}_{\mu\nu} = \epsilon_{\mu\nu\rho\sigma} F^{\rho\sigma}$ the dual field-strength tensor,
is odd under parity, and thus a pseudoscalar; in pure Yang-Mills theory, a field correlator involving this anomalous term presents a pole at the mass of the pseudoscalar glueball.

Because the $\eta_{\rm singlet}$ particle should also appear there, the following approximation has been proposed~\cite{Rosenzweig:1981cu} for an effective meson Lagrangian treatment:
\begin{equation} \label{anomalouscurrent}
 \partial_\mu K^\mu = \tilde{G}_1 + \tilde{G}_2 + \dots
\end{equation}
substituting the anomaly by a sum over the fields associated with the creation of the singlet pseudoscalar particles, including $\eta_{\rm singlet}$ and $G_{0^{-+}}$ proportional to those in Eq.~(\ref{anomalouscurrent}) (the proportionality constants are explained in~\cite{Rosenzweig:1981cu}).
The fun observation of that work is that if the mixing matrix between $\eta_{\rm singlet}$ and $G_{0^{-+}}$ is $a_{ij}$, we have
\begin{equation} \label{currentdist}
\partial_\mu K^\mu = \sqrt{3} f_\pi (a_{11}\eta_{\rm singlet} + a_{12} G_{0^{-+}} )
\end{equation}
so that experimental production of the pseudoscalar glueball proceeds by the (presumably small?) mixing $a_{12}$  with $\eta$, $\eta'$ or by higher-twist operators. This is because the pseudoscalar operator of lowest dimension (smallest number of fields and derivatives)  built from the gluon field-tensor is indeed this $F^{\mu\nu} \tilde{F}_{\mu\nu}$ combination~\footnote{This is a different way to show that the analysis of subsection~\ref{scalartensor} below applies to the $0^{++}$ and $2^{++}$ but not to the $0^{-+}$ glueball.}.

I would imagine that Eq.~(\ref{currentdist}) will need to be extended for the additional $\eta$-like mesons that may strongly share a flavor-singlet configuration and will play a role in the analysis of the pseudoscalar spectrum in years to come.

\subsubsection{Employing exotic quantum numbers}

With three gluons one can form glueballs of exotic quantum numbers, that cannot be admixed with conventional 
$q\overline{q}$ mesons because of $J^{PC}$ conservation by the strong interactions. 
Because $q\overline{q}$ mesons carry, in terms of the relative $L$ and total $S$ an angular momentum  $J\in (|L-S|,\dots L+S)$ and discrete quantum numbers $P=(-1)^{L+1}$, $C=(-1)^{L+S}$, the following $J^{PC}$ combinations are not achievable: $0^{--}$, $(2n)^{+-}$, $(2n+1)^{-+}$.
This makes them prime candidates for experimental searches as identification of a resonance featuring them excludes it as a conventional meson; still, mixing with other configurations, saliently meson-meson molecules, is still possible.

The $\eta_-$--like $0^{--}$ glueball has been a subject of contemption among QCD sum rule practitionners with Pimikov {\it et al.}~\cite{Pimikov:2017bkk} placing it at an unassailable $7\pm 1$ GeV while Qiao and Tang~\cite{Qiao:2014vva} put it at $3.8\pm 0.1$, in line with other three--gluon states~\cite{LlanesEstrada:2005jf}.

A small overview of the masses of other glueballs with exotic quantum numbers, including lattice and sum rule computations~\cite{Qiao:2015iea} suggests that a $0^{+-}$ glueball can be found in the 4.5--5 GeV region; and a $2^{+-}$ in the 4--4.3 GeV one (with the sum rule assigning it instead a much higher mass).

Searches for these objects would require multiparticle, exclusive identification in the charmonium region. 
For example, in analogy with discoveries in the $J/\psi\pi\pi$ spectrum, that showcases salient meson states such as the $1^{++}\ \chi_{c1}'(3872)$ and $1^{--}\ \psi(4260)$ mesons, attention could be given to 
$J/\psi 4\pi$, 
that couples to $0^{--}$ quantum numbers; the glueball would be detectable below the $J/\psi f_1(1285)$ if its mass is indeed as in~\cite{Qiao:2015iea}.

Since none of these glueballs is expected to populate the energy region below 3 GeV, I will not discuss them any further.

\newpage

\subsection{Production of the light scalar glueball}

Scalar mesons can be produced in multiple collision channels such as $pp$ and $p\bar{p}$, 
but for the glueballs expected below 2 GeV, a most interesting alley is the radiative $J/\psi$ decay. Because both $c\bar{c}$ quarks are annihilated in ground state charmonium decays, leaving only light quarks (that do not directly couple to charm) and radiation ({\it e.g.} gluons) behind, $J/\psi$ decays have traditionally been considered a gluon-rich environment where to look for glueballs~\cite{Close:1996yc}. Therefore, we concentrate on this channel here, though some eventual comments are found in other parts of this review.

\subsubsection{$J/\psi$ radiative decays}

Radiative decays $J/\psi\to \gamma+G$ are particularly interesting because the photon carries away the $1^{--}$ quantum numbers of the $J/\psi$, exposing the $PC=++$ glueballs with spin 0 or 2, computed to be the lightest, and other $f_0$, $f_2$ mesons. A typical such spectrum will be shown later in figure~\ref{fig:spectrumdistort}. Meanwhile, let us quickly review a typical analysis~\cite{Guo:2020akt}.

The radiative decay widths have been computed in lattice gauge theory~\cite{Chen:2014iua}, that find, approximately, the following branching fractions ($X_i=\Gamma_{J\psi\to i}/\Gamma_{J/\psi \rm total}$)
\begin{equation}
X_{\gamma G(0^{++})} \simeq 0.004(1)\ \  \ 
X_{\gamma G(2^{++})} \simeq 0.011(2) \ .
\end{equation}
These are not negligible branchings, if we compare them to 
$X_{\gamma \rm hadrons}=0.088(11) \simeq X_{\gamma gg} $ in the interpretation of the particle data group~\cite{Tanabashi:2018oca}. The lattice computation would entail that one in six radiative $J/\psi$ decays would produce a glueball; and it is supported by earlier sum-rule computations~~\cite{Narison:1996fm} that also produced $X_{\gamma G(0^{++})} \simeq 0.004-0.005$. 

Because~\cite{Close:1996yc,Guo:2020akt} $X_{\gamma f_0} \propto X_{\gamma gg} \frac{m_{f_0}\Gamma_{f_0\to gg}}{m^2_{J/\psi}}$, with known proportionality factors, the $f_0$-to glue branching fractions
$b_i := \Gamma_{f_0^i\to gg}/\Gamma_{f_0^i}$ have been reconstructed by Guo {\it et al.}~\cite{Guo:2020akt}
to be $b_{1370}=0.28(22)$, $b_{1500}=0.17(8)$ and $b_{1710}=0.85(16)$, 
in agreement with the bottom chart of figure~\ref{fig:mixing} in which the glueball configuration is dominant in the heaviest of these three mesons and does contribute a small part of the wavefunction of the other two, particularly the lightest one.

In all, this is one of the findings that drives the building consensus~\cite{Cheng:2015iaa} around most of the scalar glueball strength being found in $f_0(1710)$:
production of this meson is much stronger than that of the $f_0(1500)$ in $J/\psi$ radiative decays.

\begin{figure}[h]
\includegraphics[width=0.9\columnwidth]{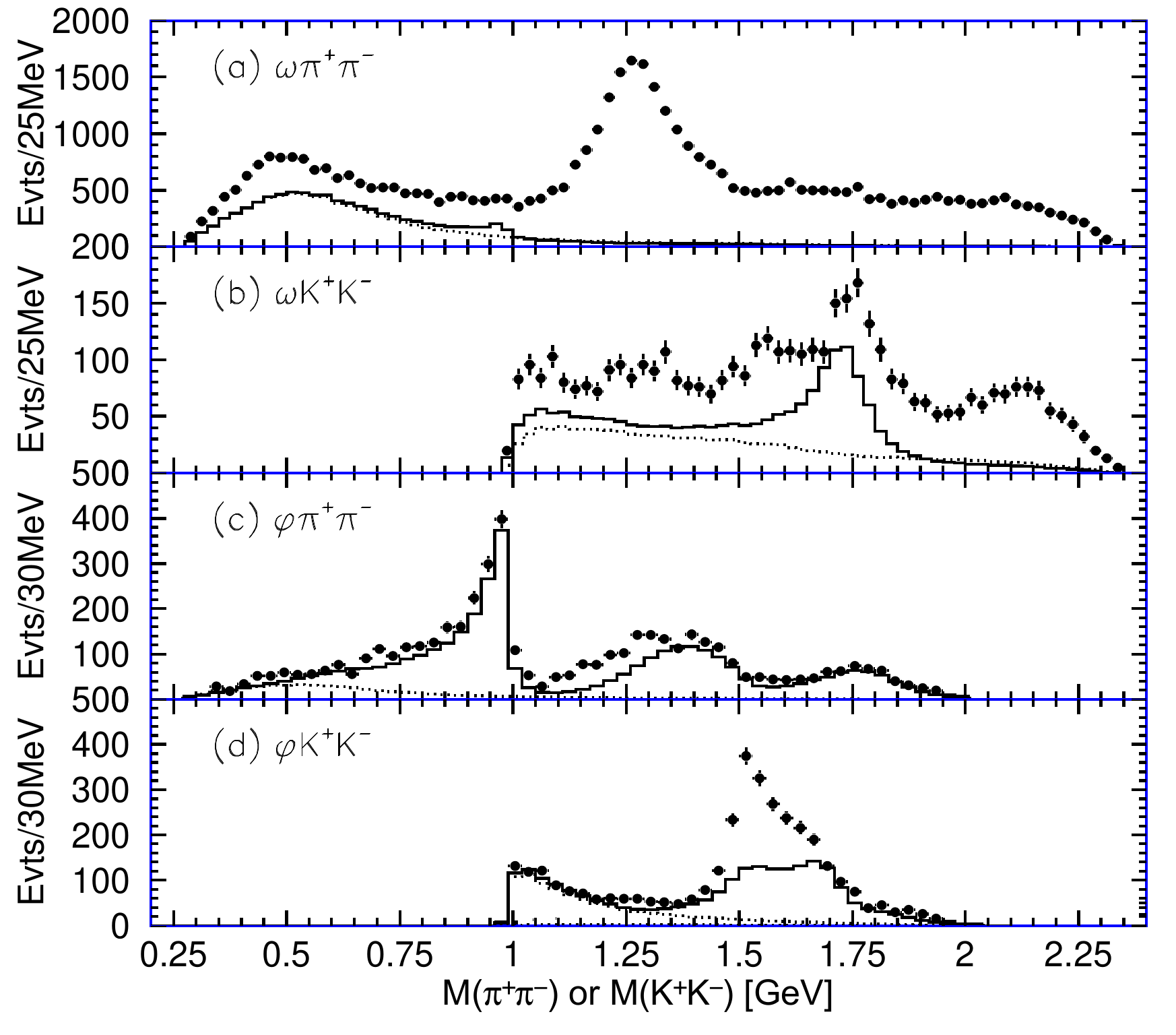}
\caption{\label{fig:MMchannels} $0^{++}$/$2^{++}$ meson spectrum from $J/\psi \to V+MM$ where the vector meson is the strong-force analog to a $\gamma$. Note the several $f_0$, $f_2$ mesons produced.
Reproduced from~\cite{Li:2006ni}, courtesy of the BES-II collaboration and of Elsevier under STM permissions guidelines.
(\emph{I thank prof. Shuangshi Fang for providing the graph file and reference}).}
\end{figure}

\subsubsection{$J/\psi$ to vector + (mesons) decays}

An interesting extension of the radiative-decay idea is to substitute the photon by a vector meson
with equal $J^{PC}=1^{--}$ quantum numbers; recoiling against that vector is the system of interest, 
often two pions or two kaons, that carries $0^{++}$ or $2^{++}$ quantum numbers. The large statistics at BES-III allow such exclusive reconstruction, shown in figure~\ref{fig:MMchannels}. Moreover, there are partial-wave analysis of various meson-meson final states that confirm $f_J$ spins as listed.

The branching fractions are not negligible: $\omega\pi\pi$ and $\omega K\bar{K}$ make up about 1\% of all $J/\psi$ decays, with $\phi \pi\pi$ and $\phi K\bar{K}$ another half a percent. Because the $\omega$ and $\phi$ are narrow and easily reconstructible, they allow access to a clean recoiling spectrum, as seen in the figure.

The figure shows that the $f_0(1710)$ is rather produced recoiling against an $\omega$ than a $\phi$ and preferentially decays to $K\bar{K}$ over $\pi\pi$. 
The broad $f_0(1370)$ bump, however, is seen to behave in the opposite way, decaying to $\pi\pi$ but being produced with more statistics against a $\phi$ vector meson, an effect that can be somewhat puzzling.

\section{Hints from and searches in high-energy scattering}

\subsection{The Pomeron and the odderon puzzle}

\subsubsection{The $2^{++}$ glueball in the Pomeron trajectory}

\begin{figure}
\begin{center}
\includegraphics[width=\textwidth]{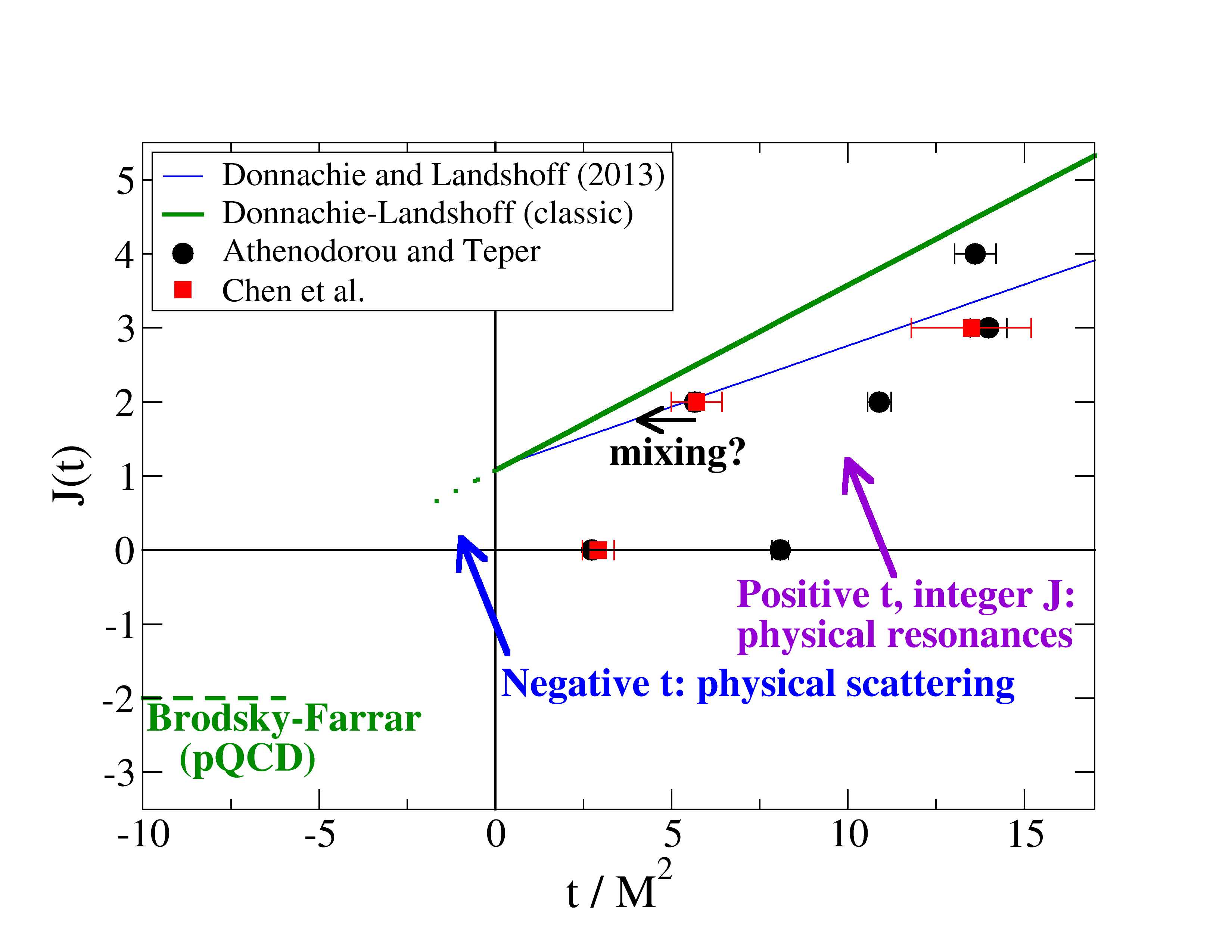}
\end{center}
\caption{\label{fig:pomeron}
Donnachie-Landshoff ``soft'' Pomeron trajectory (solid lines: higher one, the classic trajectory
from the 1990s, lower line, 2015 fit~\cite{Donnachie:2013xia}. Lattice data for $J^{++}$ glueballs 
from two different groups are represented by solid symbols. It seems clear, as has been known for long~\cite{GonzalezMestres:1979zu,Simonov:1990uq,LlanesEstrada:2000jw,Bicudo:2004tx}, that glueballs may offer an explanation of the Pomeron, and that the lightest glueball resonance that may fall near the Pomeron trajectory is the $2^{++}$ $f_2$-like glueball. Lattice data seems to put it at a mass somewhat too high, but it is possible that configuration mixing with a $q\bar{q}$ state moves the eigenvalue closer to the trajectory~\cite{Simonov:1990uq}.
}
\end{figure}

Hadron scattering amplitudes at high energies (such as $pp\to pp$ as an example) for physical $s$ and $t<0$ are known to behave as power-laws 
\begin{equation} \label{reggepower}
\sigma \propto s^{\alpha(t)-1}\ .
\end{equation} 
This functional dependence naturally arises in Regge theory~\cite{Regge:1959mz}, in which the two-body system's angular momentum $J$ is analytically continued to a complex variable $\alpha$. The function $\alpha(t)$ controls the cross-section for negative $t$, and if this variable is also continued to positive $t$ (that would correspond to the $s$ variable of $p\bar{p}$ annihilation, for example), resonances appear in the Chew-Frautschi plot shown in figure~\ref{fig:pomeron}.

The plot illustrates the leading trajectory that entails no exchange of electric charge, parity, nor charge conjugation among the scattering particles, which is due to the so called ``Pomeron'' Regge trajectory. 
Fits to $pp$ and other scattering data~\cite{Donnachie:2013xia} based on sophisticated versions of Eq.~(\ref{reggepower}) yield the discontinuous line near and to the left of the $J$ ($OY$) axis.

Far to its left on the deep $t<0$ region, pQCD predicts that elastic scattering will asymptotically follow a power law with negative exponent discussed in Eq.~(\ref{differentialcounting}) below. What is of interest for the glueball discussion is the prolongation of that first straight line to the right of the plot (solid line), where $t\to M^2>0$. 

There is no guarantee that a Regge trajectory stays linear far from the $J$ axis, as demonstrated for the $f_0(500)$~\cite{Pelaez:2015qfa}~\footnote{Incidently, the result of that work shows that this meson, popularly known as $\sigma$, is a poor glueball candidate.}. 
However, two-gluon glueballs have been computed in many model approaches~\cite{Brisudova:1997ag,LlanesEstrada:2000jw,Buisseret:2009yv,Sharov:2008zz} to fall on linear Regge trajectories $\alpha(t)=\alpha(0)+\alpha'(0) t$. 

Because two-gluon glueballs are the lightest $PC=++$ glueballs, it has long been conjectured~\cite{GonzalezMestres:1979zu,Simonov:1990uq,LlanesEstrada:2000jw,Bicudo:2004tx} that they might provide the resonances that the Pomeron trajectory produces when $\alpha_P(M^2)=J$, an integer. 
Supporting this conjecture is the fact that the slope of the Regge trajectory of $gg$ is smaller than that of quark-antiquark states, in any approach with one-gluon like color exchange. In the linearly confining potential field theory of~\cite{LlanesEstrada:2000jw}, $V\to \sigma R$ at large distance, with
\begin{equation} \label{casimir}
\frac{\sigma_{gg}}{\sigma_{q\bar{q}}} = \frac{N_c}{(N_c^2-1)/(2N_c)}
\end{equation}
the ratio of two Casimirs, yielding, in view of $\alpha' \propto \frac{1}{\sigma}$,
\begin{equation}
\frac{\alpha'_{\rm Pomeron}}{\alpha'_{q\bar{q}\ \rm Reggeon}} = \frac{4}{9}\ .
\end{equation}

Because typical Regge trajectories of conventional $q\bar{q}$ meson Reggeons have \\
$\alpha'_{q\bar{q}\ \rm Reggeon}\simeq 0.9$, if the Pomeron is identified with the $t$-channel exchange of a tower of $gg$ states with $PC=++$, its slope is predicted to be $\alpha'_{\rm Pomeron}\simeq 0.4$, in reasonable agreement with the scattering data extraction of the Pomeron by Donnachie and Landshoff~\cite{Donnachie:2013xia}. While the lattice data seems to have this higher slope, model work in Coulomb gauge QCD~\cite{LlanesEstrada:2000jw} 
is closer to the empirical Pomeron slope. 

If the excited $0^{++}$ glueball of figure~\ref{fig:glueballspectra} is ever identified, as it naturally is a radial excitation of the ground state $G$, it will allow to confirm of discard the Casimir string-tension scaling of Eq.~(\ref{casimir}). This should not be taken for granted as it is a feature of Cornell-like approaches that cast much of the confinement strength into (nonperturbative) one-gluon like exchanges with the same color factors, but there are other possibilities~\cite{Greensite:2011zz}.

Finally we can reverse the discussion and try to learn something about glueballs from high-energy Pomeron phenomenology. First of all, because the Pomeron trajectory seems to intercept the $t=0$ axis at $J=1+\epsilon$~\footnote{Technically, if $\alpha(0)=1+\epsilon$, $\sigma\propto s^\epsilon$ would violate unitarity at asymptotically high energy. While this is of no urgent concern at the LHC where the cross section of order 100 mbarn is way smaller than the $O(20)$ barn cross section of the Froissart bound, 
 some authors prefer setting $\alpha(0)=1$ exactly. Then a $J=1$ $f_1$ meson would be predicted to have zero mass, which is obviously not present in Nature. The Donnachie-Landshoff Pomeron fit nicely excludes this unwanted feature, but then unitarity needs to be corrected by multiple Pomeron exchange.}, no state with $J=0,1$ can lie on it. Therefore, the lightest and lowest-spin glueball on the Pomeron trajectory is the $2^{++}$ $f_2$-like. While Athenodorou and Teper~\cite{Athenodorou:2020ani} place its mass at $2376\pm 32$ MeV, the Pomeron would seem to prefer a mass somewhat lighter than 2.3 GeV, perhaps as low as 1.9 GeV as in the classic Jaroszkiewicz-Landshoff Pomeron $J=1.08+0.25t$ later used by Donnachie and Landshoff too.

\begin{table}
\caption{Different computations of the $2^{++}$ glueball mass, extracted from the Pomeron Regge trajectory and from various theory approaches.}
\begin{tabular}{|c|cccccc|} \hline
Method              & Pomeron  & Coulomb-$gg$ & Lattice & Constituent  & AdS-CFT & Sum rules\\ \hline
$2^{++}$ mass  & {\bf 1.9}~\cite{Jaroszkiewicz:1974ep}   & 
{\bf 2.05}~\cite{Szczepaniak:1995cw,LlanesEstrada:2000jw} & {\bf 2.38(3)}\cite{Athenodorou:2020ani}  & 
{\bf 2.59}~\cite{Simonov:1990uq} & {\bf 2.3-2.7} \cite{Vento:2017ice} & {\bf 2.0(1)}~\cite{Narison:1996fm} \\ 
(GeV) & {\bf 2.3}~\cite{Donnachie:2013xia} & 
{\bf 2.42}~\cite{Szczepaniak:2003mr} & {\bf 2.39(15)}~\cite{Chen:2005mg} & {\bf 2.53}~\cite{Buisseret:2009yv} & $\simeq${\bf 2.3}\cite{Rinaldi:2020qbm} & \\
\hline
\end{tabular}
\end{table}

\begin{figure}[h]
\includegraphics[width=5in]{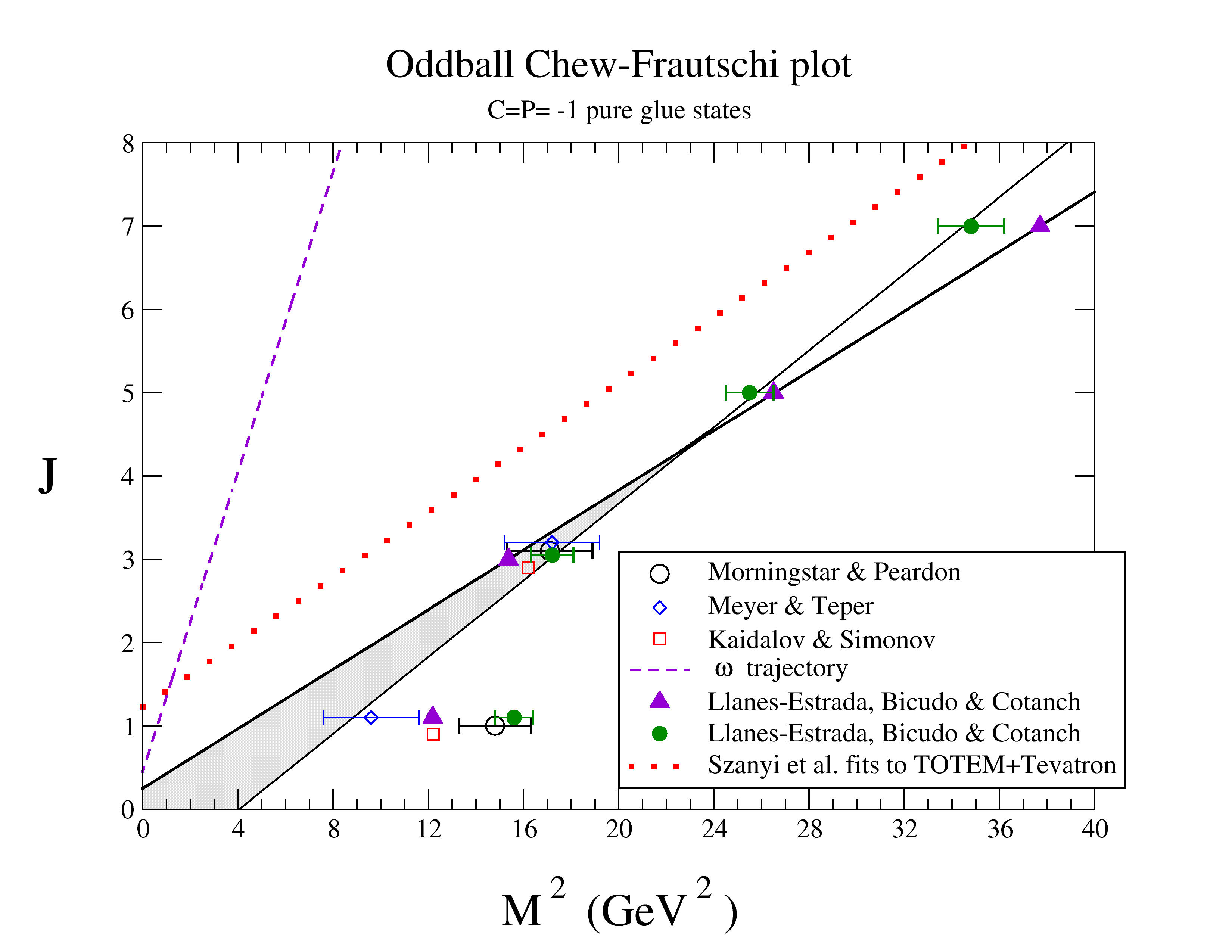}
\caption{\label{fig:odderon} Computations of the odd $C$-parity glueball spectrum (states with $J^{PC}=3^{--},5^{--},\dots$ represented by various symbols) from~\cite{LlanesEstrada:2005jf} and others quoted there lead to the conclusion that the intercept of the corresponding Regge trajectory would be $\alpha(0)<1$, as shown by the rough band reaching the $OY$ axes even below $1/2$ where conventional Regge trajectories intercept. Recent fits of high  energy scattering data~\cite{Szanyi:2019kkn} however suggest an intercept above 1 (dotted line, red online). The controversy is ongoing. 
}
\end{figure}

\subsubsection{The odderon puzzle}

\begin{figure}[h]
\includegraphics[width=\columnwidth]{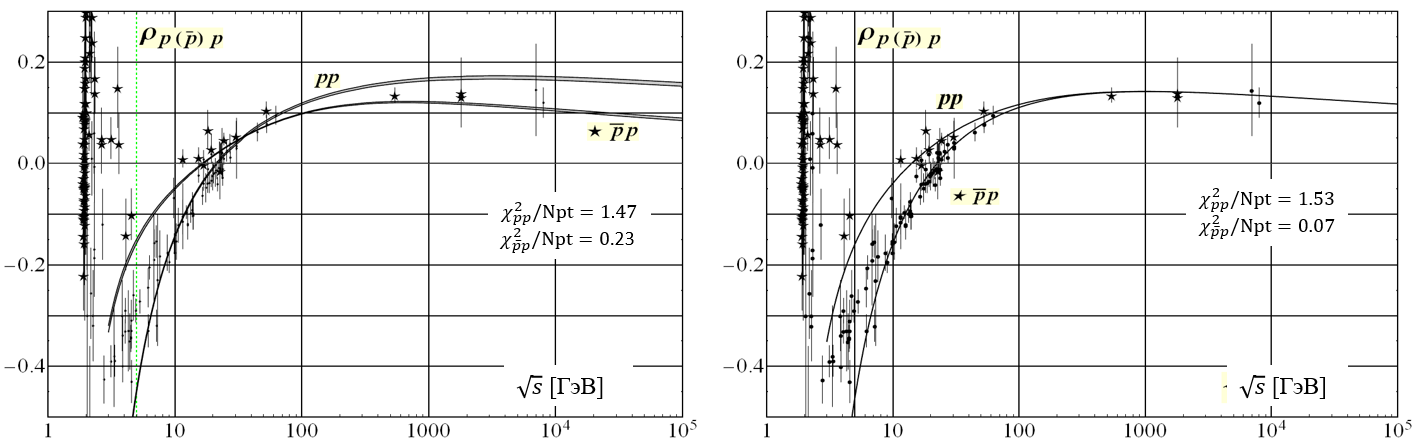}
\caption{\label{fig:pomeronodderon} According to recent work~\cite{Ezhela:2020hws,Belousov:2020rzj}, 
the total cross-section data for $pp$ and $p\bar{p}$ can be fitted with (left: $\sigma_{p\bar{p}}\neq \sigma_{pp}$) or without (right: $\sigma_{p\bar{p}}\to \sigma_{pp}$) an odderon contribution, so its existence as a crossing-odd asymptotically dominant Regge trajectory is not firmly established. Its confirmation would cause an important puzzle in our understanding of ``oddballs'' (negative $C$-parity glueballs). Figure courtesy of V. Petrov and collaborators~\cite{Ezhela:2020hws}. }
\end{figure}

Moving on, I would like to discuss the very latest developments. 
Fits to high energy data comparing the $pp$ and $p\bar{p}$ cross sections have lead to a revival of the concept of the odderon, a Regge trajectory that would give a different asymptotic cross section to the two processes. 
Such fits~\cite{Szanyi:2019kkn,Csorgo:2020rlb} seem to suggest an odderon trajectory $\alpha(t) = (1.23+0.19{\rm GeV}^{-2} t)$ with a 1.23 intercept at $t=0$ that is clearly larger than one~\footnote{Strictly speaking, because of the known asymptotic behavior, Szanyi {\it et al.}~\cite{Szanyi:2019kkn} parameterize the odderon trajectory (I have rounded off for clarity)  as
$\alpha(t) = (1.23+0.19{\rm GeV}^{-2} t)/ (1+0.032(\sqrt{t_0-t}-\sqrt{t_0})) $.
The denominator, for $t\sim 9 {\rm GeV}^2$ in the region where glueballs are important is a small $O(5\%)$ correction so we can ignore it; its importance resides, for physical $t$, in the TeV region covered by the LHC.
}. 

Earlier expectations based on computations of the odd $C$-parity glueball spectrum~\cite{LlanesEstrada:2005jf}~\footnote{The well-known work by Bartels, Lipatov and Bacca~\cite{Bartels:1999yt} deals with the BFKL-type odderon with different kinematics, as the Bjorken limit is needed in addition to high energies, and is not relevant for the glueball discussion.}
confirmed by~\cite{Kaidalov:2005kz,Cardoso:2008sb}
suggested that the Odderon Regge trajectory would not exist, because its trajectory would fall even below the conventional $\omega$ meson Regge trajectory (see Fig.~\ref{fig:odderon}), and $1-\sigma_{p\bar{p}}/\sigma_{pp}$ would be suppressed at high energy.

Other researchers~\cite{Donnachie:2019ciz,Ezhela:2020hws,Belousov:2020rzj}, analyzing the same database,
do not seem to find conclusive evidence of an odderon contribution (see figure~\ref{fig:pomeronodderon}), and it seems that more data is needed to close the discussion in this energy range. Its importance lies in that the finding of the odderon would undermine our understanding of the Pomeron as a correlated two-gluon exchange with physical resonances for integer $J$ and $t=M^2>0$, and close a window to glueballs. On the other hand, if no odderon contribution is necessary, a prediction of the whole field stands.

\subsection{Absence of Glueballs in Heavy Ion collisions?}

Hadron spectroscopy in heavy ion collisions offers interesting possibilities for identifying and classifying certain hadrons~\cite{Cho:2017dcy}. Among them, the case of Yang-Mills glueballs is, according to a part of the literature very easy: if a hadron is reconstructed in a heavy-ion collision, it is very likely not a glueball, because these ``evaporate'' or disappear from the spectrum~\cite{YepezMartinez:2012rf} very quickly at the phase transition. 
This insight was obtained, in a truncation of Coulomb gauge Yang-Mills theory, by obtaining a variational approximation to $\Omega(k)$, a screened {\it in medio} gluon-self energy minimizing the free energy at finite temperature $\delta \mathcal{F} / \delta \Omega = 0$. Thermodynamic magnitudes can then be represented in terms of that $\Omega$, for example the energy density counting glueballs 
\begin{equation}
\epsilon_1 = \frac{T^2}{V} \frac{\partial (-\beta \mathcal{F})}{\partial T}
\end{equation}
quickly above the phase transition overshoots the rigorous thermodynamical limit of Stefan-Boltzmann, 
whereas that counting individual gluons
\begin{equation}
\epsilon_2 = 2(N_c^2-1) \int \frac{d^3q}{(2\pi)^3} \frac{\Omega(q)}{e^{\beta\Omega(q)}-1}
\end{equation}
can reproduce it, suggesting that indeed glueballs have molten at the phase transition indicated by lattice data.

Unfortunately, ``glueball-like'' ordinary hadrons tend to also be relatively broad 
structures that disappear from the spectrum, unlike {\it e.g.} $\psi$ or $\Upsilon$ $q\bar{q}$ mesons. 

Likewise, nonhadronic structures such as triangle singularities also very likely drop out of the spectrum~\cite{Abreu:2020jsl} in the thermal medium. Therefore, the lack of a signal in a heavy-ion collision analysis is far from suggestive that the corresponding state could be a glueball: the statement is that, if a signal is seen in heavy-ion collisions, (a) it is more likely a hadron~\cite{Abreu:2020jsl} than in vacuum collisions and (b) it is unlikely a glueball~\cite{YepezMartinez:2012rf}.

Other investigations however suggest that there is an intermediate temperature phase below 270 MeV where glueballs are still active degrees of freedom~\cite{Stoecker:2015zea}, in which case they could contribute to RHIC/LHC phenomenology.

\section{Where to look next?}
\subsection{Counting rules and production at Belle: $0^{++}$ and $2^{++}$ glueballs}
\label{scalartensor}

In a renormalizable theory like QCD, when all scattering scales in an exclusive process
such as $AB\to CD$ become large and proportional to the total squared cm energy $s$, 
the differential cross section satisfies the Brodsky-Farrar counting rules~\cite{Brodsky:1973kr,Matveev:1973ra} that yield a simple power-law scaling with $s$,  
\begin{equation} \label{differentialcounting}
\frac{d\sigma(AB\to CD)}{dt}  = {f(\theta_{CM})\over {s^{n_i+n_f-2}}} .
\end{equation}
The power of this observation  is that a hadron--level cross section is expressed in terms of quark-gluon level constituents:
$n_i$ and $n_f$ represent the minimum number of pointlike particles
in the initial and final states.     
This idea has been exploited to predict the scaling of form factors and various cross sections and helicity selection rules.

If orbital angular momentum is included~\cite{Amati:1968kr,Ciafaloni:1968ec,Brodsky:1974vy},
one needs to take into account the short distance suppression brought about by the centrifugal factor $r^L$ (that appears in basically any formulation of hadron structure such as nonrelativistic Schr\"odinger wavefunctions, light-front ones where the radial-like variable is $\zeta^2 = b^2_\perp x(1-x)$, or Bethe-Salpeter bound-state amplitudes).
This increases the suppression of amplitudes involving a hadron with $L$ units of internal angular momentum by a factor $\left(\sqrt{s}\right)^{-L}$~\cite{Brodsky:1981kj}, with the cross sections then dropping an additional $s^{-L}$, that is, after summing all internal orbital angular momenta,
\begin{equation} \label{counting}
\frac{d\sigma}{dt}  = {f(\theta_{CM})\over {s^{n_i+n_f+L -2}}} 
\end{equation}
(at fixed angle so that $t\propto s$).

This counting rule has recently been proposed~\cite{Brodsky:2018snc,Llanes-Estrada:2018omz}
to aid with the identification of the scalar glueball among the $f_0$ states.
For the glueballs with $J^{PC}=0^{++}$, the minimum Fock space component is
$\arrowvert \vec{g}\cdot \vec{g} \rangle$ with antialigned gluon spins and no orbital angular momentum. Therefore $n_f+L=2$. 
This happens to be the \emph{slowest} falloff among all the Fock space components that can contribute to the quark-gluon Fock expansion of a scalar meson: a few are shown
in table~\ref{tab:suppression} adapted from~\cite{Brodsky:2018snc}.

\begin{table}
\caption{Power of $s$ in the QCD counting rules that suppress the production of the lowest wavefunctions in a meson Fock expansion \emph{relative to the $s$-wave glueball one} in large momentum transfer reactions involving an $f_0$ or $f_2$ meson. Introducing additional particles obviously further depresses the cross section. The glueball happens to be the most readily produced meson at high energy and momentum transfer. This is a good test to isolate the gluonium components in $0^{++}$ and $2^{++}$ mesons. \label{tab:suppression}}
\begin{center}
\begin{tabular}{|c|cccc|} \hline
Wavefunction& $gg\ar_{L=0}$  & $q\bar{q}\arrowvert_{L=1}$ & $q\bar{q}g$ & $q\bar{q} q\bar{q}$ \\
$n_f+L$     &  2    &  3         & 3         & 4 \\               
Suppression &  1    & $s^{-1}$   & $s^{-1}$  & $s^{-2}$ \\
\hline 
\end{tabular}
\end{center}
\end{table}

Because  conventional $0^{++}$ $q\bar{q}$ mesons require a p-wave, their high-energy exclusive production is suppressed respect to the gluonium $gg$. 
The same observation holds for $2^{++}$ quantum numbers: both $L=0$ $\ar gg \rangle $ glueballs compete in production experiments with $L=1$ $\ar q\bar{q} \rangle$ conventional mesons, with the glueballs dominating at high energy.  On the contrary, the $\eta$--like $0^{-+}$ glueball is an $L=1$ state competing with $L=0$  $\ar q\bar{q} \rangle$ conventional mesons, and therefore the glueball production is suppressed respect to conventional quarkonium in that channel.

Many accelerator experiments could exploit that advantage of high-energy glueball production, but particularly so Belle-II, for example by means of the reaction $e^-e^+\to \phi f_0$. 
Because the $\phi$ meson can be readily identified as an  $L=0$ $s\bar{s}$ state, $n_f=4$ for that quark-antiquark pair and two gluons for the glueball all with $L=0$, whereas
$n_i=2$ for the $e^-e^+$,  yielding $\frac{d\sigma}{dt}  = f(\theta) \frac{1}{s^4}$.

If all events in the Belle-II barrel detector were counted, (amounting to an integration over a fixed solid angle that excludes the forward direction, so $t$ is not suppressed respect to $s$), all scales are large and 
\begin{equation} 
\sigma\arrowvert_{\rm barrel} = 4\ar {\bf p}_\phi\arrowvert \arrowvert {\bf p}_{f_0}\arrowvert 
\times \int_{0}^{\cos\theta_{\rm min}}  \! d\cos\theta \ \ \frac{d\sigma}{dt}
\end{equation}
brings in one more power of $s$, resulting in the asymptotic power-law behaviors

\begin{eqnarray}\label{Glueballscaling}
\sigma \left(f_{0/2}=\ar {\bf{gg}} \ra_{L=0} +\dots \right) & \sim & \frac{\rm constant}{\bf s^3} \\ \nonumber
  & \phantom{\sim} & \\ \nonumber
\sigma \left(f_{0/2}=\ar {\bf{q\bar{q}}}\ra_{L=1} +\dots \right)       & \sim& \frac{\rm constant}{\bf s^4}   \\ \nonumber
  & \phantom{\sim} & \\ \nonumber
\sigma \left(f_{0/2}=\ar {\bf{q\bar{q}q\bar{q}}}\ra_{s-{\rm wave}} +\dots \right)       & \sim& \frac{\rm constant}{\bf s^5}   
\end{eqnarray}

If Belle-II took data {\it e.g.} at 9 and 11 GeV (off--resonance), the ratio of the cross sections at the two energies would fall by a factor, 
$ \frac{\sigma(9{\rm GeV})}{\sigma(11{\rm GeV})} \simeq 3.4 \ (gg)\ ; \ 5 \ (q\bar{q})_{L=1}\ ; \ 
7.5\ (qq\bar{q}\bar{q})$, etc. that depended on the inner structure of the scalar (eventually, tensor) meson. The large energy of this reaction entails fast separation of the two $\phi$ and $f$ mesons reducing final state interactions.

\begin{figure}[t]
\includegraphics*[width=0.48\columnwidth]{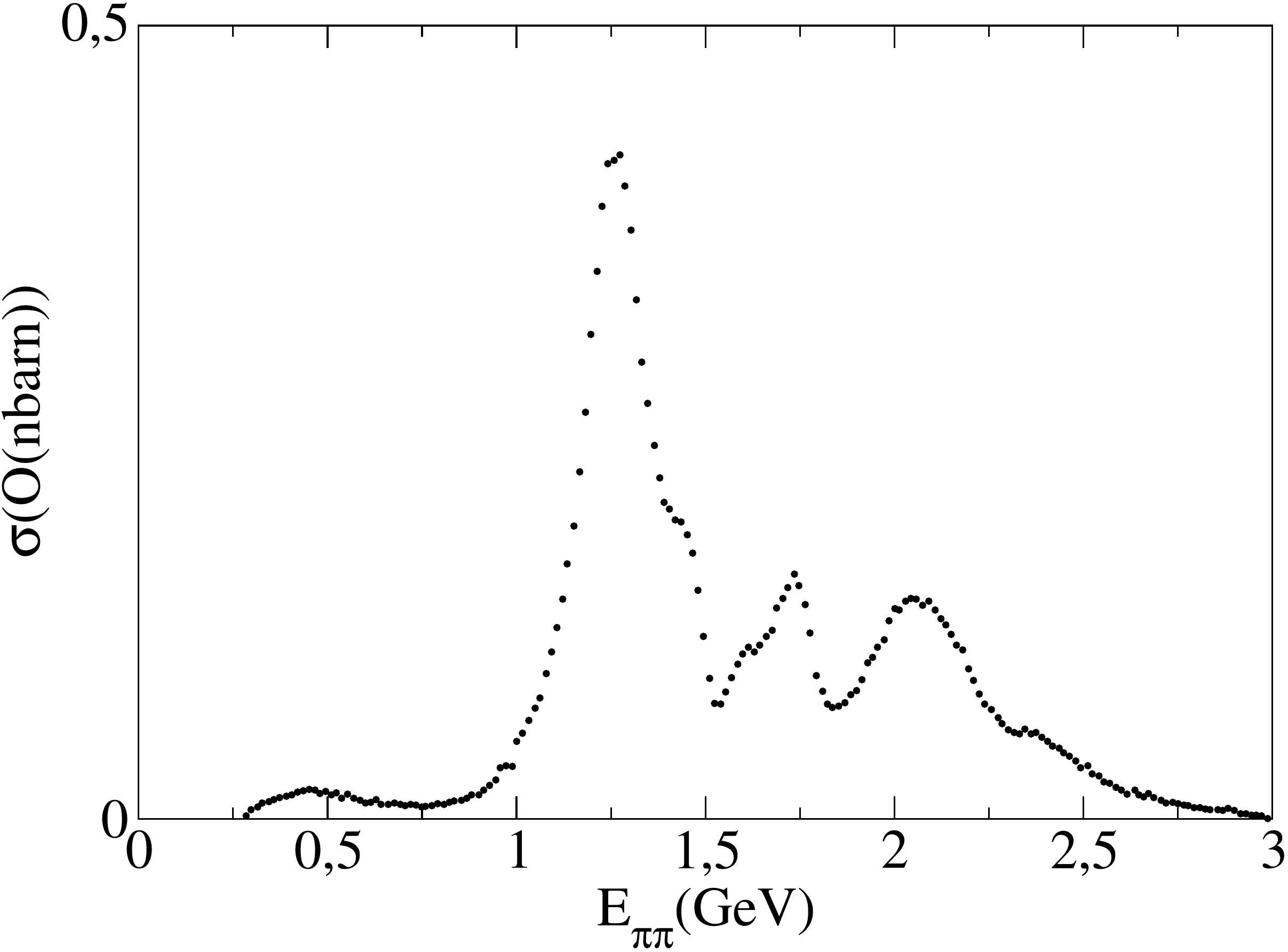}\ \ 
\includegraphics*[width=0.48\columnwidth]{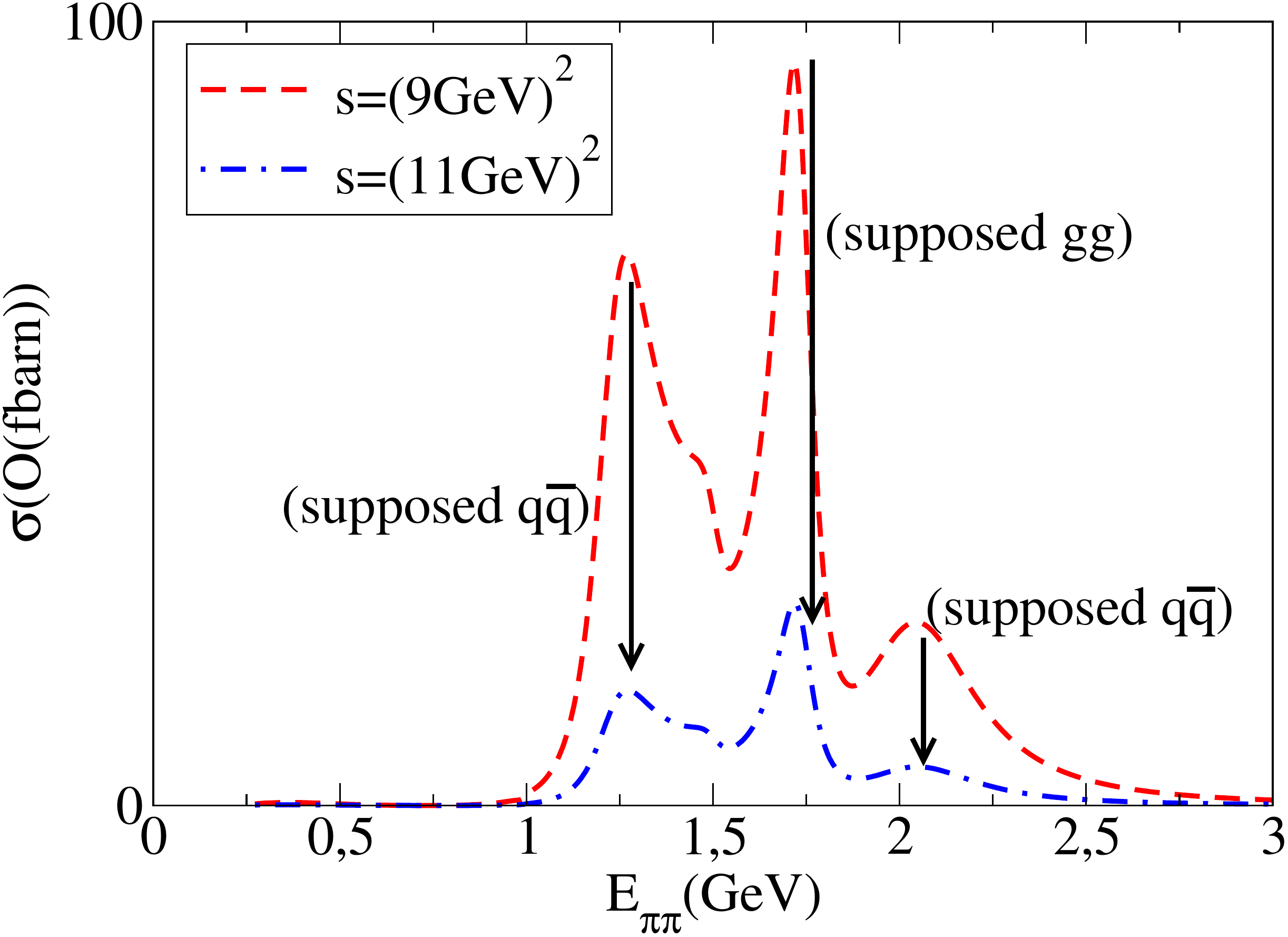}
\caption{\label{fig:spectrumdistort} 
Left: Experimental $\pi\pi$ spectrum~\cite{Bennett:2014fgt} from $J/\psi \gamma \pi\pi$.
Right: example $\pi\pi$ spectrum resulting from  $e^-e^+\to \phi f_J$ with $E= 9$ and 11 GeV, assuming  
that $f_0(1710)$ is the glueball and with absolute normalization taken from~\cite{Brodsky:2018snc}.
Whichever state dropped least in this plot upon having experimental data at hand would fit the role of the glueball. Reprinted from~\cite{Brodsky:2018snc} (Elsevier) under STM permissions guidelines.}
\end{figure}

The well known $C=+1$ $\pi\pi$ spectrum from radiative $J/\psi$ decays~\cite{Bennett:2014fgt} is shown in the left plot of figure~\ref{fig:spectrumdistort}. The typical scale here is thus at the charmonium's 3.1 GeV. 

The right plot in fig.~\ref{fig:spectrumdistort} then assumes, to exemplify, that $f_0(1710)$ is mostly the glueball and the remaining $C=+1$ states present, saliently the $f_2(1270)$, have cross sections scaling as  $q\bar{q}$ mesons. 
With $\sigma(9{\rm GeV})\sim 70$ fbarn,  70000 $\phi$--recoiling $f_0(1710)$s could be obtained at Belle-II with 1 ab$^{-1}$ of integrated luminosity (several weeks of off-resonance data), and  about 20000 events at 11 GeV, numbers that allow a check of the scaling law even after allowing for experimental cuts.

The experimental data itself, once collected, can inform the collaboration whether the energy achieved is high enough to be in the asymptotic limit $s\sim t\to \infty$, because it can test as follows whether the hadron is still behaving as pointlike without its constituents being exposed.

Profiting from the reasonable Vector Meson Dominance model, where the $\gamma$ fluctuates to a vector meson (such as $\phi(1680)$ or $Y(2175)$) and constructing an interaction Lagrangian along the lines of~\cite{Black:2006mn}, 
\begin{equation}
{\mathcal L_{\phi'\phi f_0}} = \frac{\beta}{2}f_0(\phi'_{\nu,\mu}-\phi'_{\mu,\nu})(\phi^{\mu,\nu}-\phi^{\nu,\mu}) + \frac{e}{3} \tilde{g} f_\pi^2 A^\mu \phi'_\mu\ , 
\end{equation}
hadrons behave as pointlike objects, and the prediction for the cross section is much softer than 
Eq.~(\ref{Glueballscaling}), since $n_i+n_f+L-2=2+2+0-2=2$ (as the initial state contains $e^-e^+$ and the final state two pointlike mesons). 

Thus, up to logarithms, and while the softest drop in $\sigma$ that QCD supports in Eq.~(\ref{Glueballscaling}) at large $s$ is $1/s^3$, with unstructured $\phi$ and $f_0$ the cross--section falls as
\begin{equation} \label{eq:hadroncross}
\sigma_{\rm hadron}(e^-e^+\to \phi f_0) \propto \frac{1}{s}\ .
\end{equation}
This behavior provides the experimental null hypothesis (no access to the meson's internal structure): as long as the cross section drops following the $1/s$ behavior of Eq.~(\ref{eq:hadroncross}), production is still low--energy, probing the hadron as a whole. 
Only once $\sigma$ drops as $1/s^3$ or faster can one access the intrinsic QCD counting.

\subsection{Multibeam analysis to search for the $0^{-+}$ glueball (and other $\eta$-like mesons)}

Additionally to its interest for the axial anomaly commented on in subsection~\ref{subsubsec:anomaly}, the pseudoscalar glueball is sensitive to the three-gluon scattering kernel $V^{\mu\rho\sigma}$ that extends the three-gluon vertex of pQCD to the nonperturbative, strong-coupling regime~\cite{Souza:2019ylx}, so that finding out its mass would immediately constrain the integrated strength of that function of the gluon momenta, of interest for Dyson-Schwinger studies.
 
But the spectrum of pseudoscalar, isospin-singlet mesons in the relevant mass region, around and above 2 GeV, is much less understood that the scalar one in the one and a half GeV mass range, though
there are several $\eta$-like pseudoscalar mesons below 2 GeV. The $\eta$ and $\eta'$, mixed and influenced by the anomaly as they seem to be, are clearly markers of and presumably seeded by the flavor-nonet (octet+singlet) characteristic representation of $q\bar{q}$ mesons. 

The next three possible states are $\eta(1295)$, $\eta(1405)$ and $\eta(1475)$.
The lightest, $\eta(1295)$ is almost degenerate with the $\pi(1300)$ which would suggest an ideally mixed $(u\bar{u}+d\bar{d})/\sqrt{2}$ configuration, with the $s\bar{s}$ remainder at higher mass (see minireview in \cite{Zyla:2020zbs}), and all corresponding to a radially excited quark-model nonet. Which one is that additional $\eta$ meson that would complete the nonet is more disputed.

The proposal~\cite{Albaladejo:2010tj,Liang:2013yta} that $\eta(1475)$ can be explained as a molecular-type state of composition $\eta K\bar{K}$, as they find strong binding in this channel (but not in $\eta'K\bar{K}$) would leave the lighter $\eta(1405)$ as the other largely $q\bar{q}$ state. However, the dominant decays of the higher $\eta(1475)$ state matching those of $s\bar{s}$ suggest that it is the middle one that is a supernumerary, and since the 80s its study was pursued as a possible glueball candidate (see {\it e.g.}~\cite{Masoni:2006rz}). But its mass does not match the lattice gauge theory predictions for the pseudoscalar glueball mass, that put it in the 2 GeV region, nor that of several other approaches (such as the Coulomb-gauge computations cited that require to pay the energy cost of a $p$-wave, the AdS-CFT conjecture that would make it nearly degenerate with the $2^{++}$ state, and others). 
Even more, some authors~\cite{Wu:2012pg} interpret the evidence as there being only one pseudoscalar state instead of two: this would be the traditional $\eta(1440)$ and there would be no supernumerary state in this mass region.

At higher energy yet, there could be two broad structures, the $\eta(1760)$ (with $\Gamma\sim O(250)$ MeV) and the $\eta(2225)$  (with $\Gamma\sim O(200)$ MeV). Whether any of these two, particularly the higher one, have anything to do with the pseudoscalar glueball remains to be seen. 

To produce pseudoscalar mesons in this mass range, $J/\psi$ radiative decays are not a good tool, since  $J^{PC}$ conservation in an $s$-wave decay, $1^{--}\to 0^{-+} + 1^{+-}$ cannot be exploited, as there is no $1^{+-}$ meson below 1 GeV to leave enough phase space for the high $\eta$ spectrum. The $\psi(3686)$ decays are not promising either because, though $h_1(1170)$ is light enough to leave the needed phase space, it is very broad, difficulting the multibody reconstruction. 

Belle II could profit from its higher center of mass energy and attempt the analysis of the decay chain
\begin{equation}
\Upsilon(4S) (1^{--}) \to X(0^{-+}) + h_c(1P)
\end{equation}
in which the pseudoscalar $X$, maybe not reconstructed but with spectrum obtained by the recoiling mass technique~\cite{Pakhlov:2009nj}, measuring the rest of the reaction, would contain the glueball and any other mesons in that energy range. The narrow $1P$ charmonium state with $\Gamma\simeq 0.7$ MeV, and the ample phase space, would work in favor of the search. 
Reconstructing the $h_c$ however is not so straightforward, because its dominant decay mode $\gamma \eta_c(1S)\to \gamma \eta/\eta' \pi\pi$ has to confront the 30 MeV-broad $\eta_c$.

The other promising alley is to use a $e^-e^+$ machine as a photon-photon collider, 
\begin{equation}
e^-e^+ \to e^-e^+ + {\gamma\gamma\to \rm hadrons}\ .
\end{equation}
The quantum number combinations that appear in this reaction with the leptons tagged are of course similar to those of the glueball spectrum: $0^{++}$, $2^++$ ($s$-wave), $0^{-+}$ (p-wave) ...

Belle-II could certainly dedicate some effort to the identification of pseudoscalar mesons~\cite{Shwartz:2019zle} in the 2-3 GeV energy range. In fact, the earlier Belle collaboration carried out a fruitful spectroscopy experimental program based on two-photon physics, though more focused on charmonia~\cite{Uehara:2006cj}. Their copious statistics would allow them to produce $p$-wave states of the $\gamma\gamma$ system, though not the $\eta$-glueball directly (as it is uncharged); but they could also employ the large scalar samples to search for two-pseudoscalar mesons, one of them being a $\pi^0$, by $\gamma\gamma\to G(0^{-+})+\pi^0$ as proposed early on by Wakely and Carlson~\cite{Wakely:1991ej}.

Finally, an additional, less immediate possibility, would be to rig one of the two existing $e^-e^+$ colliders with polarized beams, an upgrade that seems to be under consideration for Belle-II~\cite{Roney:2019til}.
Having polarized beams of enough purity would hopefully allow to overcome the overwhelming one-photon $e^-e^+$ annihilation background, which has $J^{PC}=1^{--}$ quantum numbers, exposing the two-photon annihilation reaction $ e^- e^+\ar_{S=0} \to 0^{-+} $.

The combination of these three production methods at lepton machines, together with $p\bar{p}$ annihilation by the PANDA experiment~\cite{Boca:2015oza,Brambilla:2014jmp,Belias:2020zwx} or at Glue-X in Jefferson Lab~\cite{Gutsche:2016wix}, that will allow production and careful study of $\eta$-like mesons above 1.9 GeV, irrespective of their components being or not charged as the proton and antiproton can annihilate via the strong force, will facilitate mixing studies such as those in subsection~\ref{subsec:symmetrymixing}. This will hopefully allow the identification of the pseudoscalar glueball in a not too distant future. As seen in table~\ref{tab:pseudoscalar}, the $\eta$-glueball will be suppressed in channels preferentially producing charged final states.

\begin{table}
\caption{\label{tab:pseudoscalar} A multibeam analysis combining data from different measurements will be essential to eventually separate the pseudoscalar Yang-Mills glueball from other $\eta$-like mesons in the 2 GeV energy region. Because gluons are uncharged, direct production in lepton machines is forbidden unless another hadron populates the final state.}
\begin{tabular}{|c|c|c|c|c|c|} \hline
Reaction                       & $p\bar{p}\to 0^{-+}$  & $\gamma\gamma \to 0^{-+}$ & $\gamma \gamma \to 0^{-+}0^{-+}$ & (polarized) $e^-e^+\to 0^{-+}$ & $\Upsilon\to 0^{-+}+h_c$\\ \hline
$q\bar{q}$, $q\bar{q}g$ \dots  &  \checkmark  & \checkmark  & \checkmark & \checkmark & \checkmark \\
Glueball & \checkmark & {\large $\times$} & \checkmark & {\large $\times$} & \checkmark\\ \hline
\end{tabular}
\end{table}

\section{Conclusions}
Gluonium or glueballs are a doubtlessly attractive piece of physics: a dense, self-bound matter-like 
state made of pure radiation, without fermions seeding it. Nature has so far not offered us another example of this configuration~\footnote{Graviton-graviton scattering is a very long shot~\cite{Blas:2020dyg}.}.

Glueballs have been searched for, and not unmistakeably identified, for over four decades. Nevertheless, searching for them has been and remains an inspiring quest to understand hadrons and is worth carrying on because the data obtained and analysis methods employed are some of the activities keeping hadron physics fascinating.

This search for Ithaca seems to at least have found a coast. Many colleagues concur that a large part of the scalar glueball configuration, in spite of mixing, is to be found in $f_0(1710)$, as most of the experimental puzzles can be resolved~\cite{Cheng:2015iaa}. Therefore,
$f_0(1370)$ and $f_0(1500)$ may have a small part of glueball component, but are, presumably, $q\overline{q}$ quarkonia to a large extent. Under this hypothesis, that  $f_0(1710)$ meson should be  a starting point for studies of the conformal anomaly, how the QCD scale arises from the Yang-Mills sector.

The two excited states almost certainly below 3 GeV and thus relevant for light quarks are the $2^{++}$ and $0^{-+}$ glueballs. They will be hidden among a largely unknown spectrum of $f_2$ and $\eta$-like mesons, respectively. There is a window to the $2^{++}$ glueball in the Pomeron Regge trajectory, on which the glueball-Pomeron conjecture reasonably maintains that it is the most prominent resonance. However, recent claims that an asymptotic odderon has been detected have cast doubt in the picture, because oddball (odd-$C$ glueballs, by contemporary usage, not hybrid mesons) computations  would predict a subleading odderon-like trajectory, that is, asymptotically equal $\sigma_{pp}$ and $\sigma_{\bar{p}p}$.

Both this glueball and the pseudoscalar one, that might open a window to glance at the axial anomaly, 
are expected to be broader than the ground state scalar one, by the many hadron channels open (no wavefunction suppression in the final state), with the flux tube model predicting $\Gamma \propto M$, so their widths would possibly be in the $O(250-350)$ MeV rather than $O(100-200)$ MeV. Finding them might amount to clarifying the entire spectrum in that mass region, just as with the $0^{++}$ one.

Additionally, the pseudoscalar glueball does not have such easily identifiable decay channels as the positive parity ones. Therefore, a promising detection strategy is by the recoil mass technique, identifying, for example, a primary $\pi_0$ meson against which the glueball may recoil. Because other $\eta$-like  mesons can behave in the same manner, a multibeam analysis comparing the same spectrum sourced from different initial states will help in sorting out which of the states had electric charge (and thus, quarks) in their configurations. 

For the $0^{++}$ and $2^{++}$ glueballs, we can additionally exploit, in future experiments, the fact that they are produced at the lowest twist in pQCD, because their leading Fock-space expansion is made of only two particles with no relative orbital angular momentum, so that they dominate over other mesons of the same quantum numbers in asymptotic production. Belle-II could study $f_0$ and $f_2$ exclusive production against, for example, a $\phi$ or similar meson, and from the cross-section falloff help identify which mesons have the largest glueball component. 

Gluonium will doubtlessly remain an object of study for years to come.

\section*{Funding  acknowledgment}

This publication is supported by EU Horizon 2020 research and innovation programme, STRONG-2020 project, under grant agreement No 824093; grants MINECO:FPA2016-75654-C2-1-P, MICINN: PID2019-108655GB-I00,  PID2019-106080GB-C21 (Spain); Universidad Complutense de Madrid under research group 910309 and the IPARCOS institute.


\end{document}